\documentclass[gsifonts,twoside,12pt]{gsipaper}
\usepackage{a4wide,gsiindex,helvet}
\usepackage[english]{babel}
\usepackage{amsmath,amsfonts,amssymb,verbatim,graphicx,subfigure,float,epsfig,
psfrag,mathrsfs}
\begin{document}

\title{\Large Phenomenology of AdS/QCD and Its Gravity Dual}

\date{\today}

\abstract{We construct the dilaton potential in the gravity dual
theory of AdS/QCD for the warp factor of Refs.~\cite{Pirner:2009gr,
Nian:2009mw}. Using this $AdS_5$-metric with properties similar to
QCD, we find that the gravity dual leads to a meaningful gauge
coupling in the region between the charmonium and bottonium mass, but
differs slightly from QCD in the extreme UV. When we fix the
ultraviolet behavior in accord with the $\beta$-function, we can
obtain good agreement with the overall heavy quark-antiquark
potential. Although the leading order proportional to
$-\frac{\alpha^{4/3}}{r}$ differs from perturbative QCD, the full
potential agrees quite well with the short distance QCD potential in
NNLO.}

\author{B. Galow$^a$, E. Meg\'{\i}as$^b$, J. Nian$^b$, H. J. Pirner$^{ab}$}

\address{$\phantom{!}^{a}$ Max Planck Institute for Nuclear Physics,
Heidelberg, Germany\\$\phantom{!}^{b}$ Institute for Theoretical Physics, University of Heidelberg, Germany}

\maketitle
\thispagestyle{empty}

\section{Introduction}
\label{sec:intro}
Maldacena's conjecture~\cite{Maldacena:1997re} states that, in the low energy
limit, the large-$N_c$ $\mathcal{N}=4$ super Yang-Mills field theory in
four-dimensional space is equivalent to the type $IIB$ string theory in
$AdS_5\times \mathrm{S}^5\,$-$\,$space. This may provide a possibility to solve
the longstanding problem of strongly coupled QCD in the low energy limit. But
a rigorous top-down approach is still far from giving any experimental
prediction. Using the idea of holography, bottom-up models can quantitatively
reproduce many results of QCD in the low energy limit gained by other existing
methods like lattice QCD. The core of these bottom-up models is to find a
reasonable non-conformal metric of the $AdS_5$-space, which incorporates
relevant physical information.

To obtain such a metric, one can either maintain the basic form of the
conformal $AdS_5$-metric and introduce a factor by hand, or assume a very
general form of the metric, and calculate its explicit parameters.
In this paper, we further investigate the warp factor given in
Refs.~\cite{Pirner:2009gr,Nian:2009mw} based on the interpretation of the fifth-dimension $z$ as a coordinate proportional to the inverse energy resolution. 
The warp factor of Refs.~\cite{Pirner:2009gr,Nian:2009mw} 
resembles the running coupling in QCD - having a
strong inverse logarithmic growth in the infrared. The Minkowskian form of the
metric with the warp factor is
\begin{equation}
ds^2=h(z)\frac{1}{(\Lambda z)^2}(-dt^2+d\vec{x}^2+dz^2),
\label{string_frame_metric}
\end{equation}
with
\begin{equation}
h(z)=\frac{\log\left (\frac{1}{\epsilon} \right )}{\log\left
[\frac{1}{(\Lambda z)^2+\epsilon}\right ]}. \label{eq:hzPirner}
\end{equation}
This metric with asymptotically conformal symmetry in the UV and infrared
slavery in the IR region yields a good fit to the heavy $Q\bar Q$-potential with
\begin{eqnarray}
\Lambda&=&264\,\text{MeV}\,, \label{eq:Lambda}\\
\epsilon&=&\Lambda^2 l_s^2\,=\,0.48\,. \label{eq:epsilon}
\end{eqnarray}

In Ref.~\cite{Nian:2009mw}, we have used this metric to calculate the
expectation value of one circular Wilson loop $\langle
\mathrm{W}\rangle$ and the correlator of two concentric circular
Wilson loops $\langle \mathrm{WW}\rangle$ from the Nambu-Goto action
of the form:
\begin{equation}
S_{NG}=\frac{1}{2\pi l_s^2}\int d^2\xi\sqrt{\det h_{ab}}\,, \label{eq:NG1} 
\end{equation}
where $l_s$ is the string length defined above. The induced world-sheet metric
in the Nambu-Goto action is called $h_{ab}$:
\begin{equation}
h_{ab}=G_{\mu\nu}\frac{\partial X^\mu}{\partial\xi^a}
\frac{\partial X^\nu}{\partial\xi^b} \,.
\end{equation}

This calculation gives reasonable results for one and two
Wilson loops. In addition to the confinement physics, we have derived
the gluon condensate at zero temperature.  The above metric,
Eq.~(\ref{string_frame_metric}), is given in the string frame, and all
the calculations mentioned above are done in this frame. Low energy
string theory can be approximated further by a gravity action with a
background scalar field, the dilaton $\phi$. When the conformal
symmetry is broken, the gravity action contains a dilaton potential
$V(\phi)$, which is no longer constant and can be calculated from the
metric with the help of the Einstein equations. Relating the energy
scale of the gauge theory living on the boundary of the $AdS_5$-space
and the bulk $z$-coordinate, we can predict the running gauge coupling
inside the region between the charmonium and the bottonium mass.

When we check whether this solution of the running coupling is
compatible with the behavior of the QCD coupling in the far
ultraviolet, we find small deviations. The reason is that our metric
in spite of many good phenomenological features has conformal
invariance in the ultraviolet. By modifying the dilaton potential
$V(\phi)$, we can make the potential consistent with the QCD
$\beta$-function and the heavy $Q\bar{Q}$-potential. We find good
agreement with the short distance behavior of the $Q\bar{Q}$-potential. Finally we close the circle of investigation and recalculate the warp factor $\bar h(z)$ of the modified metric in string frame. 

The outline of the paper is as follows: After the introduction in
Section~\ref{sec:intro}, Section~\ref{gravity} describes the formal construction
of the dilaton potential from the metric in the string frame. Section~\ref{sec:soldilaton} shows how the energy scale of QCD is related to the bulk coordinate
$z$ and gives the numerical calculation of the dilaton potential. In
Section~\ref{modification}, the resulting QCD running coupling is derived and the ultraviolet behavior of the potential is improved. Section~\ref{calculation_after_modification} applies the UV-improved dilaton potential to a calculation of the heavy quark potential and the glueball spectrum.  We give in Section~\ref{sec:dilaton-runningcoupling} a discussion about the modified dilaton potential, the running coupling and the new warp factor. Finally, Section~\ref{sec:apA} gives a final discussion and our conclusions. In Appendix~A the infrared and ultraviolet properties of the 5-dim Nambu-Goto theory with the metric of Eqs.~(\ref{string_frame_metric}) and (\ref{eq:hzPirner}) is investigated. In Appendix~B we derive the infrared properties of the UV-improved dilaton potential. In Appendix~C we show technical details for the analytical computation of the heavy $Q\bar{Q}$-potential in the small distance regime.

\section{Construction of the Gravity Dual Theory with Dilaton Potential}
\label{gravity}

We argued in the previous section that the warp factor given by
Eq.~(\ref{string_frame_metric}) proposed in Ref.~\cite{Pirner:2009gr}
produces reasonable results, as shown in Refs.~\cite{Pirner:2009gr,Nian:2009mw}. But the metric has to be consistent with gravity. In
the low energy limit, string theory can be approximated by its gravity
dual theory. In a top-down approach one obtains the $IIA/IIB$
effective action for the long-range fields of $D_3$-branes:
\begin{equation}
  S_{10D-Gravity}=\frac{1}{2\kappa_{10}^2} \int d^{10}x
\sqrt{-G^s}\left[e^{-2\phi}\left(R+4(\nabla\phi)^2 \right)-\frac{1}{2\cdot 5!}
F_5^2 \right]\,.
\end{equation}
Here $R$ is the Ricci-scalar of the gravitational field, $\phi$ is the dilaton
field and $F_5$ is the 5-form flux originating from the branes. This string
frame action is characterized by the exponential dilaton dependence in front of
the curvature scalar. In order to keep the theory simple, we neglect the axion
field $a$ and other eventual space filling branes related to quark dynamics.
After inserting the equation of motion for $F_5$ back into the action, we
transform the action from the string frame to the Einstein frame, in which the
Einstein term has the conventional form by a Weyl rescaling of the metric:
\begin{equation}
G_{\mu\nu}^E(X)=e^{-\frac{4}{3}\phi}G_{\mu\nu}^s(X)\,.
\label{Einstein_metric}
\end{equation}
Integrating over the $\mathrm{S}_5$-space and combining several terms into a
dilaton potential $V(\phi)$, we obtain the following five-dimensional action in
the Einstein frame \cite{Gursoy:2007cb}:
\begin{equation}
S_{5D-Gravity}=\frac{1}{2\kappa_5^2}\int d^5x \sqrt{-G^E} \left (R
-\frac{4}{3}\partial_{\mu}\phi\partial^{\mu}\phi-V(\phi)\right)\,.
\label{5D_action}
\end{equation}
In principle, there can be an additional Gibbons-Hawking term in the
action \cite{York:1972sj,Gibbons:1976ue}. But since this term does not
contribute to the variation with respect to the metric $G_{\mu\nu}^E$,
it does not affect the Einstein equations, which determine all the
physical quantities at zero temperature. We must emphasize that, at
finite temperature, only Einstein equations themselves are not enough
to determine all the physical quantities, but the action is also
important, hence at finite temperature the Gibbons-Hawking term
contributes to the thermodynamic quantities. In this paper, we only
focus on the $T=0$ case, therefore, we can neglect the Gibbons-Hawking
term.

We assume that the background dilaton potential denoted by $V(\phi)$
incorporates some information of the 5-form $F_5$ and higher curvature
corrections extending the range of applicability. The background field
$\phi(z)$ has a $z$-dependence reflecting necessary corrections at
higher energies. Let us comment on the dimensionalities of the
quantities introduced in the action, Eq.~(\ref{5D_action}). The Ricci
scalar $R$ has dimension $[R]=\frac{1}{\mbox{length}^2}$, $\kappa_5^2$
has dimension $[\kappa_5^2]=\mbox{length}^3$.  Consequently, the bulk
field $\phi$ is dimensionless, and the dimension of the dilaton
potential $V(\phi)$ is $[V(\phi)]=\frac{1}{\mbox{length}^{2}}$. Our
space-time metric has Minkowski signature with the sign convention
$(-,+,+,+,+)$. We emphasize that all quantities in
Eq.~(\ref{5D_action}) have to be in the Einstein frame. In the
remaining part of this paper, we simply write $G_{\mu\nu}$, instead of
$G_{\mu\nu}^E$. After variation of the Einstein-frame action,
Eq.~(\ref{5D_action}), with respect to the metric $G_{\mu\nu}$, one
obtains the following equations of motion
\begin{equation}
(\underbrace{R_{\mu\nu}-\frac{1}{2}\,R\,G_{\mu\nu}}_{\equiv
E_{\mu\nu}})-\underbrace{(\frac{4}{3}\partial_{\mu}\phi\partial_{\nu}\phi-\frac{
1}{2}G_{\mu\nu}(\frac{4}{3}\partial_{\sigma}\phi\partial^{\sigma}\phi
+V(\phi)))}_{\equiv T_{\mu\nu}}=0\,.
\label{grav_eom}
\end{equation}

The Einstein equations contain the energy momentum tensor $T_{\mu\nu}$,
\begin{equation}
E_{\mu\nu}=T_{\mu\nu}.
\label{gravity_eom}
\end{equation}
With the above chosen signature of the metric and assuming just a flat metric,
we obtain $T_{00}=\frac{2}{3}(\phi')^2+\frac{1}{2}V(\phi)$, where
the prime denotes derivative with respect to $z$. Up to a normalization factor
of the kinetic term, this agrees with the definition of the energy of a scalar
field.~\footnote{Note in Refs.~\cite{Gursoy:2007cb,Gursoy:2007er} $V(\phi)$ is
defined as the negative of our potential.}

The warped metric in the string frame, Eq.~(\ref{string_frame_metric}),
becomes the Einstein-frame metric $G_{\mu\nu}$ through
Eq.~(\ref{Einstein_metric}):

\begin{equation}
ds_E^2=e^{-\frac{4}{3}\phi}h(z)\frac{1}{(\Lambda
z)^2}(-dt^2+d\vec{x}^2+dz^2).
\label{Einstein_metric_5D_1}
\end{equation}

The resulting Einstein equations, Eqs.~(\ref{grav_eom}), can be solved for the
dilaton. Besides the Einstein equations, the action given
by Eq.~(\ref{5D_action}) yields another Euler-Lagrange equation, which can be
obtained by varying the action with respect to $\phi$:
\begin{equation}
\Box\phi-\frac{3}{8}\frac{dV(\phi)}{d\phi}=0.
\label{dilaton_eom}
\end{equation}
This equation contains only redundant information and is not independent of the
Einstein equations, Eqs.~(\ref{grav_eom}). Hence Eq.~(\ref{dilaton_eom}) needs
no further treatment. This is a consequence of the fact that $\phi$ is
actually not an independent field, but part of $G_{\mu\nu}$.

Now we turn to the Einstein equations, Eqs.~(\ref{grav_eom}). To be
consistent with Refs.~\cite{Gursoy:2007cb,Gursoy:2007er}, we use the following
form of the space-time metric in the Einstein frame
\begin{equation}
G_{\mu\nu}=e^{2A(z)}\cdot\mbox{diag}(-1,1,1,1,1).
\label{conformal_flat_metric}
\end{equation}
Comparing this metric with Eq.~(\ref{Einstein_metric_5D_1}), we have

\begin{equation}
e^{2A(z)}=e^{-\frac{4}{3}\phi}h(z)\frac{1}{(\Lambda z)^2}.
\label{relation1}
\end{equation}

We first calculate the tensor $E_{\mu\nu}$ in terms of $A(z)$ and its
derivatives. Then we compute the components of the energy-momentum tensor 
$T_{\mu\nu}$ under the assumption that $\phi=\phi(z)$. 
\begin{equation}
-T_{11}=T_{22}=T_{33}=T_{44}=-\frac{2}{3}(\phi')^2-\frac{1}{2}e^{2A(z)}V(\phi),
\end{equation}
\begin{equation}
T_{55}=\frac{2}{3}(\phi')^2-\frac{1}{2}e^{2A(z)}V(\phi).
\end{equation}
Thus, we end up with  only two independent equations of motion, namely
\begin{eqnarray}
3((A'(z))^2+A''(z))=-\frac{2}{3}(\phi')^2-\frac{1}{2}e^{2A(z)}V(\phi),
\label{phi_prime}\\
6(A'(z))^2=\frac{2}{3}(\phi')^2-\frac{1}{2}e^{2A(z)}V(\phi).
\label{V_phi}
\end{eqnarray}
Adding these equations we obtain a formal expression for the dilaton potential:
\begin{equation}
V(\phi(z))=-e^{-2A(z)} (9(A'(z))^2+3A''(z)).
\label{EEOM1}
\end{equation}
Multiplying Eq.~(\ref{phi_prime}) by $(-1)$ and then adding it to
Eq.~(\ref{V_phi}) we find an important relation between the dilaton and
the metric profile:
\begin{equation}
(\phi')^2=\frac{9}{4}((A'(z))^2-A''(z)).
\label{EEOM2}
\end{equation}
These two equations, Eq.~(\ref{EEOM1}) and Eq.~(\ref{EEOM2}), are identical with
Eq.~(2.16) in Ref.~\cite{Gursoy:2007cb}, up to a different sign of the dilaton
potential. The different sign stems from the fact that we have a
minus sign in front of the potential in the action, Eq.~(\ref{5D_action}).
We have chosen the potential in this way, because we want the $T_{00}$-component
of the energy-momentum tensor to be the sum of the kinetic and the potential energy.
We see that Eq.~(\ref{EEOM2}) depends on the profile $A(z)$, which is a function
of the warp factor $h(z)$ and the dilaton field $\phi(z)$ (cf.
Eq.~(\ref{relation1})). The resulting second order differential equation for 
$\phi(z)$ needs two boundary conditions, which we will obtain from the QCD running
coupling constant once the bulk coordinate $z$ is connected with the energy
scale $E(z)$.

\section{The Energy Scale $E(z)$ and the Solution for the Dilaton Potential}
\label{sec:soldilaton}

Before we turn to the numerical solution of Eq.~(\ref{EEOM2}), we must do some
preparations. Perturbative string theory is based on a topological expansion in
the string coupling $g_s$, which generates factors  $e^{-\chi\phi}$ in the
string partition function, where $\chi$ is the Euler characteristic of the
string surface, e.g. $\chi=2$  for a sphere. Therefore the string coupling is
proportional to~$e^\phi$. We choose the proportionality constant equal to unity, i.e.
\begin{equation}
  g_s = e^{\phi} \,.
\end{equation}
The elastic amplitude of two closed strings (glueballs) of order
$\mathcal{O}(g_s^2)$ corresponds to a contribution
$\mathcal{O}(\alpha^2)$ in the boundary field
theory. In general, the $AdS/CFT$ correspondence relates the string
coupling $g_s$ of perturbative string theory to the Yang-Mills
coupling $\alpha$ in the following manner:
\begin{equation}
  g_s=\frac{g_\text{\tiny YM}^2}{4\pi}=\alpha \,.
\end{equation}
In order to simplify the notation, in the following we will use $\alpha$ to denote both the Yang-Mills coupling related to the string coupling and the QCD running coupling.

The warping of the bulk space relates the bulk coordinate $z$ to the
energy scale $E(z)$ associated with $z$ via the gravitational
blue-shift. It is pedagogical to use the radial coordinate $r \propto
1/z$. In $SU(N)$, one has $N$ $D_3$-branes as gravitational source
located at $r\equiv\frac{L^2}{z}\to 0$,\footnote{In our notation the
radius of the $AdS_5$-space is $L = \frac{1}{\Lambda}$, see
Eq.~(\ref{string_frame_metric}).} $E_{r\to\infty}$ denotes the value of
the energy scale at the holographic boundary $r\to\infty$, when $E_r$
is the value of the energy at an arbitrary value of $r$. The
blue-shift is given by the dimensionless ratio
\begin{equation}
\frac{E_r}{E_{r\to\infty}}=\sqrt{\frac{G_{tt}(r\to\infty)}{G_{tt}(r)}},
\end{equation}
where $G_{tt}$ denotes the temporal component of the metric. In the limit
$r\to\infty$, we are far away from the branes, where the space-time is 
asymptotically flat, which yields $G_{tt}(r\to\infty)=-1$. Hence, the
blue-shift reads
\begin{equation}
E_{r\to\infty}=E_r\sqrt{-G_{tt}(r)}\quad\text{or
equivalently}\quad E_{r\to\infty}=E_z\sqrt{-G_{tt}(z)}.
\end{equation}

In the unmodified $AdS_5$-space, $G_{tt}=-\frac{1}{(\Lambda z)^2}$ and hence
$E_{r\to\infty}=E_z \frac{1}{\Lambda z}$. For $z\to0$, we find the UV regime of
the boundary field theory. This is in agreement with the intuition
underlying the renormalization group interpretation of the $z$-coordinate,
which was instrumental to guess the warp factor $h(z)$. For $E_z$ one can
choose an arbitrary value of the energy scale. In order to simplify the
expression, we choose the confinement scale given by $\Lambda=264\,\text{MeV}$.
This leads to the following explicit formulas in the Einstein frame
\begin{align}
  E_{r\to\infty} & = e^{-\frac{2}{3}\phi(z)}\frac{\sqrt{h(z)}}{z}
\label{Energy_gauge_1}\\
  {} & = \alpha^{-\frac{2}{3}}\frac{\sqrt{h(z)}}{z}
\label{Energy_gauge_2}\\
  {} & = e^{A(z)}\cdot \Lambda. \label{Energy_gauge_3}
\end{align}
Suppose that we know the value of the coupling constant $\alpha$
at a given energy scale $E=E_{r\to\infty}$, then we can find the corresponding
value of $z$ from Eq.~(\ref{Energy_gauge_2}).

At a given value of $z$, $\phi(z)=\mathrm{log}(\alpha)$ gives just one
boundary condition for Eq.~(\ref{EEOM2}). In order to obtain the
second boundary condition to Eq.~(\ref{EEOM2}), we need a second value
of the QCD running coupling. It is not easy to choose two appropriate
energy values. The reason is the following. In the UV limit it is
questionable whether the $AdS/CFT$ correspondence is still valid. On
the other hand, in the IR limit, there is no reliable measurement of
the running coupling. We think that the region between the charmonium
mass and the bottonium mass, i.e. between $3\,\mathrm{GeV}$ and
$8\,\mathrm{GeV}$, is a reasonable region where the modeling of
$AdS/QCD$ with the warp factor of Eq.~(\ref{string_frame_metric})
should work well. Therefore, we propose as input values for $\alpha$
\cite{PDG}:
\begin{equation}
  \alpha(3\,\mathrm{GeV})=0.25241,\quad
  \alpha(8\,\mathrm{GeV})=0.18575.
\label{runningcouplingvalues}
\end{equation}
This also means that we have implicitly set $N_c=3$ and $N_f=4$. Although there
is no entry in our model for color and flavor, we have fitted our final result
to the Cornell potential, therefore we have implicitly made a choice for these
parameters.

\begin{figure}[!ht]
  \begin{center}
     \epsfxsize 10cm \epsffile{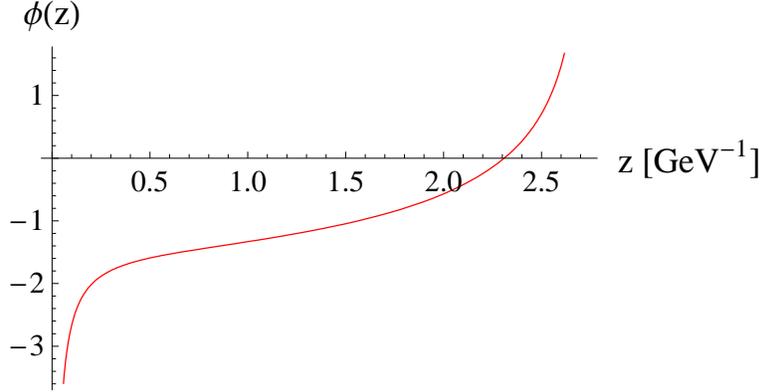}
  \end{center}
  \caption{The dilaton field profile $\phi$ as a function of the bulk coordinate~$z$, computed from the warp factor $h(z)$ of Eq.~(\ref{eq:hzPirner}).}
\label{Dilatonprofile}
\end{figure}

The fact that one has to fix two integration constants using initial conditions contrasts with the analysis of Refs.~\cite{Gursoy:2007er,Gursoy:2008za}, where the authors show that just one initial condition is enough. One possibility studied in Refs~\cite{Gursoy:2007er,Gursoy:2008za} is fixing one of the parameters by requiring that the bulk singularity is not of the ``bad kind'', which means that the singularity should be repulsive to physical fluctuations. In our case the singularity in the infrared is of the ``good kind''. As we explain in Appendix~A, the reason why we have to handle with two integration constants is that we do not use the perturbative $\beta$-function as starting point.

With the above conditions we solve Eq.~(\ref{EEOM2}) numerically, and obtain
the dilaton field $\phi$ as a function of $z$ (cf. Fig.~\ref{Dilatonprofile}).
Note the strong variations of the dilaton field for small and large $z$. For large $z$, infrared confinement at low energies is felt.

In Appendix~A we have investigated the infrared and ultraviolet properties of $\phi(z)$ analytically. Defining 
\begin{equation}
\xi = z_{\textrm {\tiny IR}} - z \,. \label{eq:xideftext}
\end{equation}
with 
\begin{equation}
z_{\textrm {\tiny IR}} = \sqrt{1-\epsilon}/\Lambda \,,
\end{equation}
we find in the infrared
\begin{equation}
\phi(\xi) = \frac{3}{16}\left(\log\frac{\xi}{\omega_{\textrm {\tiny IR}}}\right)^2 + \kappa_{\textrm{\tiny IR}} + {\cal O}(\xi\log\xi) \,,  \qquad \xi \to 0, \label{eq:phiir2text}
\end{equation} 
with
\begin{equation}
\omega_{\textrm {\tiny IR}} = 4.55 \; {\rm GeV}^{-1}  \,, \qquad \kappa_{\textrm{\tiny IR}} = -0.758 \,, \label{eq:xi0kappatext}
\end{equation}
and in the ultraviolet
\begin{equation}
\phi(z) = -\frac{\omega_{\textrm {\tiny UV}}}{z} + \kappa_{\textrm{\tiny UV}} + {\cal O}(z) \,, \qquad z\to 0 \,. \label{eq:phiuvtext}
\end{equation}

\begin{figure}[!ht]
  \begin{center}
     \epsfxsize 10cm \epsffile{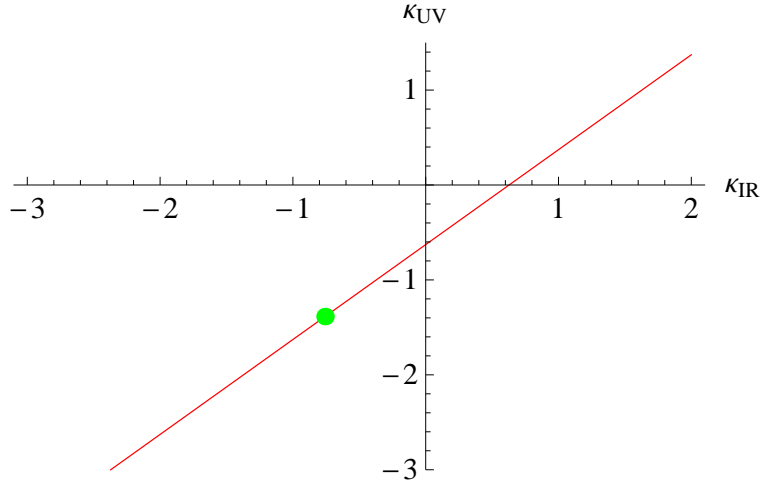}
  \end{center}
  \caption{\small The UV parameter $\kappa_{\textrm{\tiny UV}}$ as a function of the IR parameter $\kappa_{\textrm{\tiny IR}}$, cf. Eqs.~(\ref{eq:phiir2text}) and (\ref{eq:phiuvtext}), when $\omega_{\textrm {\tiny IR}}$ and $\omega_{\textrm {\tiny UV}}$ are fixed to the values quoted in Eqs.~(\ref{eq:xi0kappatext}) and (\ref{eq:z0etatext}). The point corresponds $\kappa_{\textrm{\tiny IR}}= -0.758$ and $\kappa_{\textrm{\tiny UV}}=-1.386$, which follows from the initial condition Eq.~(\ref{runningcouplingvalues}). The functional dependence is given by Eq.~(\ref{eq:etakapparelation}).}
\label{fig:etakappa}
\end{figure}

The parameters $\kappa_{\textrm{\tiny UV}}$ and $\kappa_{\textrm{\tiny IR}}$ are related, in the sense that setting $\kappa_{\textrm{\tiny UV}}$ in the UV then sets $\kappa_{\textrm{\tiny IR}}$ in the IR. $\omega_{\textrm {\tiny UV}}$ and $\omega_{\textrm {\tiny IR}}$ are related in the same way. From a numerical computation of~$\phi(z)$ in the full regime $0<z<z_{\textrm { \tiny IR}}$ we find that Eq.~(\ref{eq:xi0kappatext}), or equivalently Eq.~(\ref{runningcouplingvalues}), leads~to:
\begin{equation}
\omega_{\textrm {\tiny UV}} = 0.1285 \;{\rm GeV}^{-1} \,, \qquad \kappa_{\textrm{\tiny UV}} =  -1.386 \,. \label{eq:z0etatext}
\end{equation}
We show in Fig.~\ref{fig:etakappa} the relation between the parameters $\kappa_{\textrm{\tiny UV}}$ and $\kappa_{\textrm{\tiny IR}}$ when $\omega_{\textrm {\tiny IR}}$ and $\omega_{\textrm {\tiny UV}}$ are fixed to the values quoted in Eqs.~(\ref{eq:xi0kappatext}) and (\ref{eq:z0etatext}). The functional dependence is
\begin{equation}
\kappa_{\textrm{\tiny UV}} = \kappa_{\textrm{\tiny IR}} - 0.628 \,. \label{eq:etakapparelation}
\end{equation}

\begin{figure}[!ht]
  \begin{center}
     \epsfxsize 10cm \epsffile{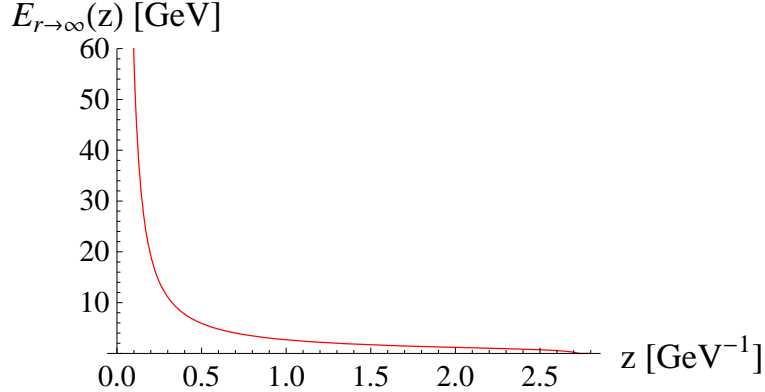}
  \end{center}
  \caption{The energy scale as a function of the bulk coordinate~$z$, corresponding to the warp factor $h(z)$ of Eq.~(\ref{eq:hzPirner}).}
\label{EYM_of_Z}
\end{figure}

Inserting $\phi(z)$ into Eq.~(\ref{Energy_gauge_1}), we calculate immediately
the associated energy scale $E_{r\to\infty}(z)$, which is shown in
Fig.~\ref{EYM_of_Z} as a function of the $z$-coordinate in the fifth dimension.
At the lower edge of the $z$-scale, we have a strongly increasing energy
$E_{r\to\infty}(z)$ (cf. Fig.~\ref{EYM_of_Z}).

\begin{figure}[!ht]
  \begin{center}
      \epsfxsize 8cm \epsffile{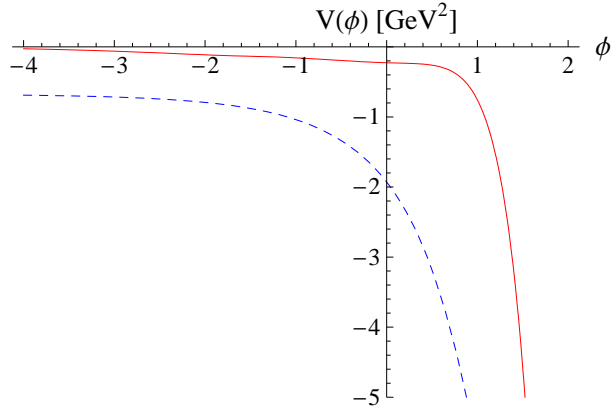}
  \end{center}
  \caption{The dilaton potential as a function of the dilaton field $\phi$.  The result corresponding to the warp factor $h(z)$ of Eq.~(\ref{eq:hzPirner}) is displayed as a full (red) line.  We show for comparison as a dashed (blue) line the modified dilaton potential given by Eq.~(\ref{analyticDilatonPotential}) with the parameters provided in Sec.~\ref{calculation_after_modification}.} 
\label{Dilatonpotential}
\end{figure}

Consequently, the $A(z)$ in the metric can be calculated from
Eq.~(\ref{Energy_gauge_3}). With $A(z)$, the other Einstein equation,
Eq.~(\ref{EEOM1}), gives us $V(z)$. Combining $V(z)$ with $\phi(z)$, we obtain
the dilaton potential $V(\phi)$, which is shown in Fig.~\ref{Dilatonpotential}.
The result shows an approximately constant dilaton potential until
$\phi(z)=1$. Beyond this point the dilaton potential falls rapidly. Recall
that in the $AdS$-space, the ``cosmological'' term is negative and slowly
varying due to the asymptotic conformal behavior of the warp factor. In the
conformal limit, $\phi'=0$, and the dilaton potential should have the value
$-\frac{12}{L^2}$, which can solve the Einstein equations, Eqs.~(\ref{EEOM1})
and (\ref{EEOM2}). 

Questions about the stability of the vacuum because of the large and
negatively unbound dilaton potential have to be analyzed, but it is
well known that due to Breitenlohner-Freedman bound negative second
order derivatives in the dilaton potential do not cause problems in
the presence of gravity \cite{Breitenlohner:1982jf,
Breitenlohner:1982bm}.

\section{Constraining the Dilaton Potential by the QCD $\beta$-Function}
\label{modification}

\begin{figure}[!ht]
  \begin{center}
      \epsfxsize 9cm \epsffile{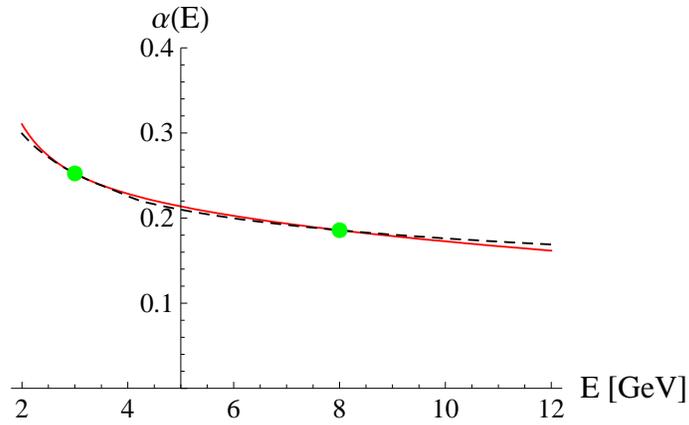}
  \end{center}
  \caption{The running coupling as a function of the energy scale. Full (red) line corresponds to the string theory result that follows from the relation 
$\alpha =e^{\phi}$, using the warp factor $h(z)$ of Eq.~(\ref{eq:hzPirner}). We show as a dashed (black) line the experimental values of the running coupling from the PDG data~\cite{PDG}. The two points correspond to the input conditions, Eq.~(\ref{runningcouplingvalues}).}
\label{Alpha-1}
\end{figure}

From the energy scale $E(z)$ and the dilaton profile $\phi(z)$, we are now able
to calculate the value of the strong coupling constant at any energy scale
$\alpha=e^{\phi(z)}$. The gravity dual of string theory allows
to interpolate the QCD coupling between our boundary values at $3\,\mathrm{GeV}$
and $8\,\mathrm{GeV}$ in a satisfactory manner, as one sees from the comparison
of $\alpha = e^{\phi}$ in Fig.~\ref{Alpha-1} with the strong
coupling from the PDG web tool \cite{PDG}. The good description of the coupling
adds another positive feature to the warp factor proposed in Refs.~\cite{Pirner:2009gr,Nian:2009mw}.

\begin{figure}[!ht]
  \begin{center}
     \epsfxsize 10cm \epsffile{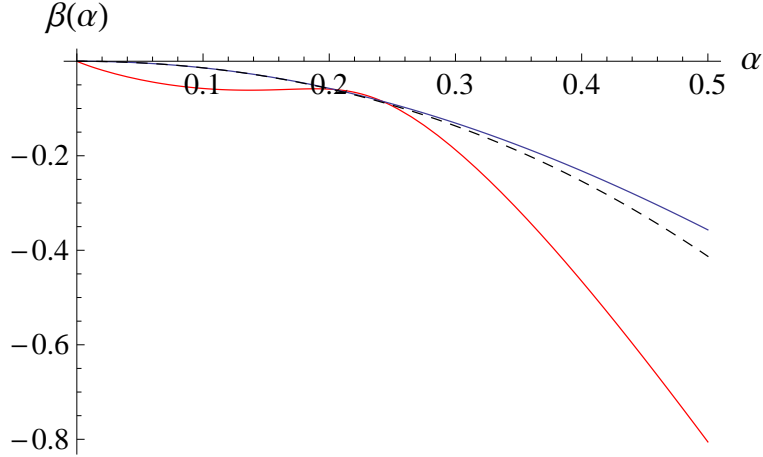}
  \end{center}
  \caption{\small The $\beta$-function as a function of the running coupling $\alpha$. We show as a full (red) line the string theory result that follows from the warp factor $h(z)$ of Eq.~(\ref{eq:hzPirner}), using Eq.~(\ref{calculatebeta}). We show for comparison as a full (blue) line the result corresponding to Eq.~(\ref{newbeta}) with the parameters provided in Sec.~\ref{calculation_after_modification}. Dashed (black) line is the QCD $\beta$-function given by Eq.~(\ref{QCDbeta}).}
\label{betafct}
\end{figure}

When we investigate the behavior of the running coupling in the deep
UV more closely, we expect some contradiction with QCD. As much as
conformal behavior is favored in correlation functions, where the
leading behavior up to logarithmic corrections is correctly
reproduced, we have to deviate from the correct running of the coupling in the
deep ultraviolet, since our metric $h(z)$ has scale independence in this
limit. We start with the definition of the $\beta$-function:
\begin{equation}
  \beta\,\equiv\,E\frac{d\alpha }{d E}.
\end{equation}
Eq.~(\ref{Energy_gauge_3}) tells us that the energy can be expressed as the
product of $e^A \cdot \Lambda$. Using $\alpha =e^\phi$, we obtain:
\begin{equation}
  \beta\,\equiv\,E\frac{d\alpha }{d
E}=e^A\Lambda\cdot\frac{d(e^\phi)}{d(e^A\Lambda)}=\frac{e^\phi
d\phi}{d A}=\frac{e^{\phi(z)}\cdot \phi'(z)}{A'(z)}.
\label{calculatebeta}
\end{equation}
All quantities in the last expression are calculable from the warp factor
$h(z)$. In Fig.~\ref{betafct} we show the $\beta$-function from our $AdS/QCD$
model together with the QCD $\beta$-function at two-loop level. In QCD the $\beta$-function has the following form~\cite{vanRitbergen:1997va}:
\begin{equation}
  \beta(\alpha)=-b_0 \alpha^2 - b_1 \alpha^3,
\label{QCDbeta}
\end{equation}
with 
\begin{equation}
b_0=\frac{1}{2\pi}\left(\frac{11}{3}N_c-\frac{2}{3}N_f\right)\,,\qquad \textrm{and} \qquad b_1=\frac{1}{8\pi^2}\left(\frac{34}{3}
N_c^2-\left(\frac{13}{3} N_c-\frac{1}{N_c}\right)N_f\right) \,.
\end{equation} 
As argued before, we have set $N_c=3$ and $N_f=4$. In this case,
$b_0=\frac{25}{6\pi}$, and $b_1=\frac{77}{12\pi^2}$. One sees that the
agreement is very good near $\alpha =0.25$, i.e. near the charmonium
mass region, where the phenomenological adjustment has been done. For
smaller values of $\alpha$ there is a sizable discrepancy.  Explicitly
in this region, the $\beta$-function stemming from the warp factor
$\propto\,-\alpha$, not $\propto\,-\alpha^2$ as in QCD. This explains
the deviation visible in Fig.~\ref{betafct} for small $\alpha$.  We
have derived in Appendix~A an analytic expression for $\beta(\alpha)$
in the deep ultraviolet, cf. Eq.~(\ref{eq:betahPirner}). The strongly
increasing warp factor $h(z)$ in the infrared leads to a stronger beta
function for large~$\alpha$.

How can one repair this problem and make the dilaton potential
consistent with QCD in the ultraviolet? The basic idea of
Refs.~\cite{Gursoy:2007cb} and \cite{Gursoy:2007er} is to use the QCD
$\beta$-function itself as a starting point, and derive the metric
from the $\beta$-function. The resulting metric is then unambiguously
consistent with the QCD $\beta$-function, as expected, and the running
coupling calculated from this new metric is necessarily correct. This
procedure presents a systematic approach to define the dilaton
potential in the ultraviolet. In the infrared, for large positive
values of the dilaton field there remains the problem to choose a
parametrization of the potential. Calculations of the string tension
have been proposed as tests of this parametrization at zero
temperature~\cite{Zeng:2008sx} or of the spatial string tension at finite
temperature \cite{Alanen:2009ej}. In our case, we will build on the
phenomenological work done in Refs.~\cite{Pirner:2009gr,Nian:2009mw} and will
fit the constrained form of the potential to the heavy quark
potential. 

In the following, we will review the important formulas given in
Refs.~\cite{Gursoy:2007cb} and \cite{Gursoy:2007er}. In the so-called
``{\it domain wall coordinates}'' $du \equiv e^A dz$:
\begin{equation}
  ds^2=e^{2A}(-dt^2+d\vec{x}^2)+du^2,
\end{equation}
the Einstein equations become
\begin{equation}
  3\ddot{A}+12\dot{A}^2=V,
\label{newEEOM1}
\end{equation}
\begin{equation}
  \ddot{A}=-\frac{4}{9}\dot{\phi}^2,
\label{newEEOM2}
\end{equation}
where the dot denotes the derivative with respect to $u$. After defining two
auxiliary variables
\begin{eqnarray}
  W &\equiv& -\frac{9}{4}\dot{A}\,,  \label{defW}  \\
  X &\equiv& -\frac{3}{4}\cdot \frac{d\mathrm{log}W}{d\phi}\,, \label{defX}
\end{eqnarray}
we may rewrite the Einstein equations using $W$:
\begin{equation}
  \dot{A}=-\frac{4}{9}W,\quad \dot{\phi}=\frac{d W}{d \phi},\quad V = 
\frac{4}{3}\left(\frac{dW}{d\phi}\right)^2 - \frac{64}{27}W^2 \,.
\label{newEinstein}
\end{equation}
$W(\phi)$ plays the role of a superpotential. Several equivalent expressions hold for $X$:
\begin{eqnarray}
  X &=&\frac{\beta(\alpha)}{3\alpha}\,, \label{exprX1} \\
  X &=&\frac{\dot{\phi}}{3\dot{A}}\,, \label{exprX2} \\
  X &=& \frac{1}{3\alpha}\cdot\frac{d\alpha}{d A}\,.  \label{exprX3}
\end{eqnarray}

When the $\beta$-function is known, we can calculate
$X$ through the first expression, Eq.~(\ref{exprX1}). With this $X$, $W$ is
obtained as a solution of Eq.~(\ref{defX}). Consequently, $\dot{A}$ and
$\dot{\phi}$ can be calculated through Eq.~(\ref{defW}) and Eq.~(\ref{exprX2}),
respectively. Finally, the general form of the dilaton potential $V$, is
determined from the last equation of the three Eqs.~(\ref{newEinstein}) as
\begin{equation}
  V(\phi)=V_0\cdot (1-X^2)\cdot e^{-\frac{8}{3}\int_{-\infty}^\phi
X(\tilde{\phi})d\tilde{\phi}}.
\label{Vform1}
\end{equation}
Therefore, for a $\beta$-function given over the whole range of
$\alpha$, the dilaton potential is fixed.  In the IR region, we do not
know the correct form of the $\beta$-function. If we want to impose
confinement in the IR region, some forms of the $\beta$-function are
excluded, but there still can be several possible classes to achieve
confinement \cite{Gursoy:2007er}. It is well known that the definition
of the $\beta$-function becomes dependent of the quantity one studies,
when the gauge coupling becomes strong. Therefore, the question
arises, which parametrization of the $\beta$-function should one
choose in the infrared. One possible choice is given by Eq.~(5.1) in
Ref.~\cite{Gursoy:2007cb}.  Here we propose another possible choice
which combines the correct UV-behavior with some integrable form,

\begin{equation}
  \beta(\alpha) = -b_2\alpha + \left[b_2\alpha  +
\left(\frac{b_2}{\bar{\alpha}}-b_0\right)\alpha^2 +
\left(\frac{b_2}{2\bar{\alpha}^2} - \frac{b_0}{\bar{\alpha}}-b_1 \right)
\alpha^3 \right] e^{-\alpha/\bar{\alpha}}\,.
\label{newbeta}
\end{equation}
This parametrization has the required $\beta$-function in the UV region as limit,
\begin{displaymath}
  \alpha \to 0:\quad \beta(\alpha )\approx
-b_0\alpha^2 - b_1\alpha^3,
\end{displaymath}
and confinement property in the IR region \cite{Gursoy:2007er}. 

We can get a constraint on the parameters $b_2$ and $\bar\alpha$ by
demanding a good behavior for the running coupling.  For $\alpha
<\bar{\alpha}$ the QCD-coupling is strictly perturbative, whereas for
$\alpha > \bar{\alpha}$ the $\beta$-function is characterized by the
non-perturbative linear term $-b_2 \alpha$.  When we consider
the coupling in the region $ 0.6 \,\textrm{GeV }< E < 15
\,\textrm{GeV}$, we obtain a good fit of the running coupling for
values $ 1.2 < b_2 < 3$ and $\bar \alpha >0.27 $ along the line~\footnote{This interval for $b_2$ will be further reduced after imposing the requirements of confinement and the infrared singularity being repulsive to physical modes, cf. Appendix~B. }

\begin{equation}
\frac{b_2}{\bar \alpha} =5.09. \label{eq:b2alpha}\\
\end{equation}
The $\chi^2/\textrm{d.o.f.}$ is very close to its minimum in the entire region of $b_2$. The perturbative running for energies larger than the charmonium 
mass is guaranteed by the limit on $\bar \alpha$. On the other hand, the $\beta$-function given by Eq.~(\ref{newbeta}) leads to a confining theory if $b_2 \ge 3/2$. To see that we need only to study the IR behavior of the corresponding function~$X(\alpha)$, cf.~Eq.~(\ref{exprX1}). In this limit
\begin{equation}
\lim_{\alpha\to\infty} \left(  X(\alpha) + \frac{1}{2}\right) \log\alpha = \lim_{\alpha\to\infty} \left(  -\frac{b_2}{3} + \frac{1}{2}\right) \log\alpha  \le 0 \Longleftrightarrow b_2 \ge \frac{3}{2}  \,, \label{eq:critconfbeta}
\end{equation} 
which constitutes the general criterion for confinement in its version of the $\beta$-function, Ref.~\cite{Gursoy:2007er}.

The dilaton potential corresponding to the parametrization of $\beta(\alpha)$ is
\begin{eqnarray}
  V(\alpha) &=& V_0 \left(1 -
\left(\frac{\beta(\alpha)}{3 \alpha}\right)^2\right) \left(\frac{\alpha}{\bar{\alpha}}\right)^\frac{8 b_2}{9} \nonumber \\
  {} && \quad\cdot \mathrm{Exp}\left[\frac{4}{9} \left( (2\gamma-3)b_2 +
4 b_0\bar{\alpha} + 2 b_1\bar{\alpha}^2  \right)\right] \nonumber \\
  {} && \quad\cdot \mathrm{Exp} \left[ \frac{4}{9} e^{-\frac{\alpha}{\bar{\alpha}}} \left( 3 b_2 - 4 b_0 \bar{\alpha} - 2
b_1\bar{\alpha}^2 + (\frac{b_2}{\bar{\alpha}}   -2 b_0 - 2 b_1
\bar{\alpha})\alpha\right) \right]   \nonumber \\
  {} && \quad\cdot \mathrm{Exp}\left( \frac{8 b_2}{9}\cdot
\mathrm{ExpIntegralE}\left[1,\frac{\alpha}{\bar{\alpha}}\right]
\right). \label{analyticDilatonPotential}
\end{eqnarray}
In this expression, $\gamma$ is the Euler's constant, and
$\mathrm{ExpIntegralE}(n,z)\equiv \int_1^\infty \frac{e^{-zt}}{t^n}
dt$ is the exponential integral function, $b_0$ and $b_1$ are the
coefficients appearing in the QCD $\beta$-function given by Eq.~(\ref{QCDbeta}), while $V_0$, $\bar{\alpha}$ and $b_2$ are undetermined constants. An interesting exercise is to expand Eq.~(\ref{analyticDilatonPotential}) in the UV. The result is
\begin{eqnarray}
V(\alpha) &=& V_0 \bigg\{ 
1+\frac{8}{9} b_0 \alpha +\frac{1}{81} \left( 23 b_0^2 + 36 b_1 \right) \alpha^2 \nonumber \\
&& \hspace{.2cm}+ \frac{2}{2187 \bar\alpha^3} \left[ 54 b_2 - 162 b_0 \bar\alpha  -324 b_1 \bar\alpha^2 +20 b_0^3 \bar\alpha^3 + 189 b_0 b_1 \bar\alpha^3\right] \alpha^3 \nonumber \\
&& \hspace{.2cm} + {\cal O}(\alpha^4) 
\bigg\}  \,.  \label{eq:Vdilaton}
\end{eqnarray}
The leading orders are determined by the UV parameters $b_0$ and
$b_1$, and the unknown constants $b_2$ and $\bar\alpha$ start
contributing at ${\cal O}(\alpha^3)$. The same feature is shared by
the parametrization of Ref.~\cite{Gursoy:2009jd}, although in this
reference the order $\alpha^3$ is replaced by $\alpha^{8/3}$. The
behavior of the dilaton potential for larger values of $\alpha$ can be phenomenologically determined by fitting to the heavy quark potential which we
will do in the next section.

\section{Fit of Parameters to $Q\bar{Q}$ Potential and Running Coupling}
\label{calculation_after_modification}

With the modified dilaton potential of
Eq.~(\ref{analyticDilatonPotential}) we can calculate the heavy
quark-antiquark potential.  In Ref.~\cite{Pirner:2009gr} the heavy
$Q\bar{Q}$-potential was in very good agreement with the Cornell
potential. Now, we recalculate it with the modified dilaton potential
$V(\phi)$ to focus on two particular questions: Firstly, how does the
potential from string/gravity theory compare with the potential from
three loop perturbation theory? This serves as a test whether the
improvements on the $\beta$-function pay off in the UV behavior of
observables.  Secondly, can the long distance string tension help us
fix the remaining parameters? In Ref.~\cite{Pirner:2009gr} the
calculation was done in the bulk $z$-coordinate, but in this section
it is better to work with the variable $\alpha =e^\phi$. A general
derivation for the heavy $Q\bar{Q}$-potential using $\alpha =e^\phi$
has been given in Ref.~\cite{Zeng:2008sx}. In the following we will
add the correct short distance and long distance analysis for the
first time. We refer to Appendix~C for further discussions and
computation in the small separation limit.

The first step is to derive the explicit form of the metric which is
consistent with the $\beta$-function given by Eq.~(\ref{newbeta}). We work
in coordinates dependent on the running coupling $\alpha$ as a
variable, instead of $z$ or $u$. From the domain wall coordinates
relation~$e^A dz = du$, one may easily derive
\begin{equation}
e^A dz = \frac{d\alpha}{\alpha \dot\phi} \,. \label{eq:eAdz}
\end{equation} 
Then using Eqs.~(\ref{defW}), (\ref{exprX1}) and (\ref{exprX2}) and solving Eq.~(\ref{defX}) we get
\begin{equation}
\frac{d\alpha}{dz} = \frac{1}{\bar\ell} e^{A-D} \,, \label{eq:dadz1}
\end{equation}
with $\bar\ell$ given by

\begin{equation}
\bar\ell \equiv \frac{6}{\sqrt{-3 V_0}} \,, \label{eq:ell}
\end{equation}
while $A$ and $e^D$ are functions of $\alpha$ given by

\begin{equation}
A(\alpha) = A_* + \int_{\alpha_*}^{\alpha}
\frac{1}{\beta(a)} da \,, \label{eq:Aalpha}
\end{equation}
and 

\begin{equation}
e^D= -\frac{1}{\beta(\alpha)} 
\exp \left[\frac{4}{3}\int_0^{\alpha}
\frac{\beta(a)}{3 a^2}da \right]\,, \label{eq:eDDD}
\end{equation}
with the fixed constants $\alpha_*$ and $A_*$ defined at the energy $E$:

\begin{equation}
\alpha_*=0.25241 \,, \qquad \qquad  E=e^{A_*}\Lambda = 3 \, \textrm{GeV} \,.\label{eq:cond1alpha}
\end{equation}
Then the new metric in Euclidean space, which includes the new warp factor $\bar h(z)$, follows from Eq.~(\ref{string_frame_metric}) using Eqs.~(\ref{relation1}) and (\ref{eq:dadz1}), and it reads

\begin{equation}
  d\bar s^2 = \bar h(z(\alpha)) \frac{1}{(\Lambda
z(\alpha))^2}(-dt^2+d\vec{x}^2)+e^{\frac{4\phi}{3}}\cdot \bar\ell^2 e^{2D}
d\alpha^2 \,, \label{eq:metricbar}
\end{equation}
with $\bar h(z)$ given by
\begin{eqnarray}
\bar h(z) &=& e^{2A(z)}e^{\frac{4}{3}\phi}(\Lambda z)^2\,. \label{relation} 
\end{eqnarray}

One advantage of the present computation starting from the $\beta$-function given by Eq.~(\ref{newbeta}) is that we need~$\alpha_*$ as single input value for $\alpha$, in contrast to the two values we needed within the formalism based on the warp factor $h(z)$ of Secs.~\ref{gravity} and \ref{sec:soldilaton}, cf. Eq.~(\ref{runningcouplingvalues}).~\footnote{With the $\beta$-function of the previous model we cannot define $e^D$ in the same way because we effectively need a cutoff at small~$\alpha$, cf.~Eqs.~(\ref{eq:eDDD}) and (\ref{eq:betahPirner}). In this case the computation of $d\alpha/dz$ should be done in a different way. Using Eqs.~(\ref{defW}), (\ref{exprX2}) and (\ref{eq:eAdz}) one has
\begin{equation}
\frac{d\alpha}{dz} = -\frac{4}{9} \beta(\alpha) W(\alpha) \,
e^{A(\alpha)} \,,
\end{equation}
where $A(\alpha)$ is given by Eq.~(\ref{eq:Aalpha}) and
\begin{equation}
W(\alpha) = W(\alpha_{\textrm {\tiny ct}}) \, e^{-\frac{4}{3}\int_{\alpha_{\textrm {\tiny ct}}}^{\alpha} \frac{\beta(a)}{3 a^2} da} \,.  \label{eq:Wapp}
\end{equation}
The cutoff $\alpha_{\textrm {\tiny ct}}$ introduces a new integration constant, which is multiplicative and related to a factor $e^{\kappa_{\textrm {\tiny UV}}}$ in $\alpha$, cf. Eq.~(\ref{eq:phiuvtext}). The correct QCD $\beta$-function in the UV, however, renders the integration in Eq.~(\ref{eq:Wapp}) finite. No multiplicative constant is needed. Only a single input value $\alpha_*$ is sufficient.} 

Note that in $A(\alpha)$ the $\beta$-function enters the integral in
the denominator, and in $e^D(\alpha)$ it appears once in the
denominator and once in the numerator of the exponential
function. Taking into account that $\beta(\alpha) < 0 $ one finds
that both $A$ and $e^D$ become large for small~$\alpha$. For large
$\alpha$, $e^D$ becomes small, while $A$ behaves
as~$-\frac{1}{b_2}\log\alpha$ in this regime.

The general procedure to compute $V_{Q\bar{Q}}$ within the
classical approximation is similar to the one used in
Refs.~\cite{Pirner:2009gr,Zeng:2008sx}. The heavy quark potential follows from the Nambu-Goto action for a rectangular Wilson loop with a short spatial side and a much longer
time side, i.e.
\begin{equation}
  \langle \mathrm{W}\rangle\simeq e^{-T\cdot V}\simeq e^{-S_{\mathrm{NG}}}\,.
\label{appCNambuGotoPotential}
\end{equation}
The picture is given by a string stretched between a quark and an antiquark, located at $x_1 = \frac{\rho}{2}$ and $x_2 = -\frac{\rho}{2}$ respectively, which dips into the bulk of the background $AdS_5$-space. The separation $\rho$ between two quarks as well as the potential energy can be expressed as functions of $\alpha_0$, which is the value of $\alpha$ at the mid-point between the quark and the antiquark.

In principle the Nambu-Goto action $S_\textrm{NG}$ can now contain a new string length $\bar l_s$ compared with the Nambu-Goto action defined in section~\ref{sec:intro},

\begin{equation}
S_\textrm{NG} = \frac{1}{2\pi\bar l_s^2 } \int d^2 \xi \sqrt{\det \bar h_{ab}} \,, \label{eq:NG2}
\end{equation}
where $\bar h_{ab}$ is the new induced worldsheet metric defined by 
$\bar h_{ab} = \bar G_{\mu\nu}\frac{\partial X^\mu}{\partial \xi^a}\frac{\partial X^\nu}{\partial \xi^b}$.

To obtain the heavy $Q\bar{Q}$-potential, we need to express the separation
$\rho$ between the quark and the antiquark, as well as the value of the
potential $V_{Q\bar{Q}}$ as functions of $\alpha_0$, which is the value of
$\alpha$ at the mid-point between the quark and the antiquark.
Similarly, $A_0 =A(\alpha_0)$ and $\phi_0=\phi(\alpha_0)$ are
the values of $A$ and $\phi$ at the mid-point. The separation between
the quark and the antiquark is given by~\cite{Zeng:2008sx}
\begin{equation}
 \rho(\alpha_0) = 2\bar\ell e^{-A_0}\cdot \int_0^{\alpha_0}
\frac{e^{D-3\tilde{A}}\cdot\tilde{\alpha}^{-\frac{4}{3}}
}{\sqrt{1-\tilde{\alpha}^{-\frac{8}{3}} e^{-4\tilde{A}}}}
d\alpha \,,
\label{finalrho}
\end{equation}
with
\begin{eqnarray}
\tilde{A} &\equiv&  A-A_0 \,, \\
\tilde{\phi} &\equiv& \phi-\phi_0 \,, \\
\tilde{\alpha}  &\equiv& \frac{\alpha }{\alpha_0} \,.
\end{eqnarray}

The bare potential $V_{Q\bar{Q}}$ calculated from the Nambu-Goto
action is divergent, so we have to regularize it. The divergence means
that the quark-antiquark pair becomes infinitely heavy. An obvious
way to remove this divergence is to subtract the rest mass of the two
heavy quarks~\cite{Zeng:2008sx, Maldacena:1998im}. Then the finite part of the
potential is
\begin{equation}
V_{Q\bar{Q}}(\alpha_0) = \frac{\bar\ell \alpha_0^{\frac{4}{3}}e^{A_0}}{\pi \bar l_s^2}
\left[\int_0^{\alpha_0} d\alpha
\frac{\tilde{\alpha}^{\frac{4}{3}}
e^{D+\tilde{A}}
\left(1-\sqrt{1-\tilde{\alpha}^{-\frac{8}{3}}
e^{-4\tilde{A}}}\right)}{\sqrt{1-\tilde{\alpha}^{-\frac{8}{3}}
e^{-4\tilde{A}}}} - \int_{\alpha_0}^\infty d\alpha
\, \tilde{\alpha}^{\frac{4}{3}}\cdot e^{D+\tilde{A}}
 \right].
\label{finalVqq}
\end{equation}
Combining Eq.~(\ref{finalrho}) with Eq.~(\ref{finalVqq}), we obtain
the heavy quark potential as a function of the separation between the
quark and the antiquark. In the numerical computation of the second
integral of Eq.~(\ref{finalVqq}) we replace the variable
$\alpha \to \frac{\alpha_0}{\hat{\alpha}}$, so that
the integral transforms into an integration between $0$ and $1$ for
the variable $\hat{\alpha}$, which is much easier to compute. The
regularization procedure ensures that the integrals are ultraviolet
convergent as can be easily proved.

In order to fix the three parameters $V_0$, $\bar\alpha$ and $b_2$ of
the dilaton potential and the string constant $\bar l_s$, we will
study separately the short distance, i.e. the ultraviolet (UV) regime, and
the large distance, i.e. the infrared (IR) regime. The parameter $V_0$ is relevant in the UV, while $\bar\alpha$ and $b_2$ become important in the IR, and $\bar l_s$ naturally scales the potential.

Let us first study the infrared properties of integrals appearing in
Eqs.~(\ref{finalrho}) and (\ref{finalVqq}). For this purpose we focus
on the large $\alpha$ behavior of the $\beta$-function, where $-b_2
\alpha $ is the relevant term, cf. Eq.~(\ref{newbeta}). As we show in Sec.~\ref{modification}, a value $b_2 \ge 3/2$ ensures that the theory is confining. On the other hand, in the limit $\alpha \to \infty$ the argument inside the square root in Eqs.~(\ref{finalrho}) and (\ref{finalVqq}) becomes
\begin{equation}
1-\tilde{\alpha}^{-\frac{8}{3}} e^{-4\tilde{A}} \simeq 1 - \tilde{\alpha}^{\frac{4}{b_2}-\frac{8}{3}} \,, \qquad  0 \le \tilde\alpha  \le 1 \,,
\end{equation} 
which is negative for~$ b_2 > 3/2$. So, a value of $b_2 > 3/2$ means
that $\alpha_0$ cannot exceed some upper limit $\alpha_0^*$, and in
the limit $\alpha_0 \to \alpha_0^*$ then $\rho$ diverges. As an
example, $\alpha_0^* = 1.38$ for $b_2 = 1.7$, and $\alpha_0^* = 1.08$
for $b_2 = 2.3$. Note that these upper limits are not too high, taking
into account that the PDG data approximately relate $\alpha =
1.45^{+0.94}_{-0.43}$ to $E=0.6\,{\textrm GeV}$ (see
Eq.~(\ref{eq:b2alpha}) and discussion).

The condition that the second integral in Eq.~(\ref{finalVqq}) is
finite upon integration to infinity
\begin{equation}
\int_{\alpha_0}^\infty d\alpha  \frac{1}{b_2}
\alpha_0^{\frac{1}{b_2}-\frac{4}{3}} \alpha^{\frac{1}{3} -
\frac{1}{b_2} - \frac{4}{9} b_2} < \infty\,, \label{eq:intIR1}
\end{equation}
necessitates $\frac{1}{3} - \frac{1}{b_2} - \frac{4}{9} b_2 <
-1$, which is fulfilled only if $b_2 \ne 3/2$ and
positive. The integration (\ref{eq:intIR1}) is convergent, but this convergence
is very slow for the values of $b_2$ we consider here. As an example,
the convergence for the value $b_2 = 2.3$ is reached only when $\alpha
\simeq 10^{20} \alpha_0$. In practice we handle this problem by
analytically computing the integral for large values of $\alpha$,
where $\beta(\alpha) \approx -b_2 \alpha $. This approach is excellent
for the interval $(5\alpha_0,\infty )$.  For values close to
$\alpha_0$, i.e. in the interval $(\alpha_0,5\alpha_0)$, we perform a
numerical integration.

\begin{figure}[!ht]
  \begin{center}
     \epsfxsize 9cm \epsffile{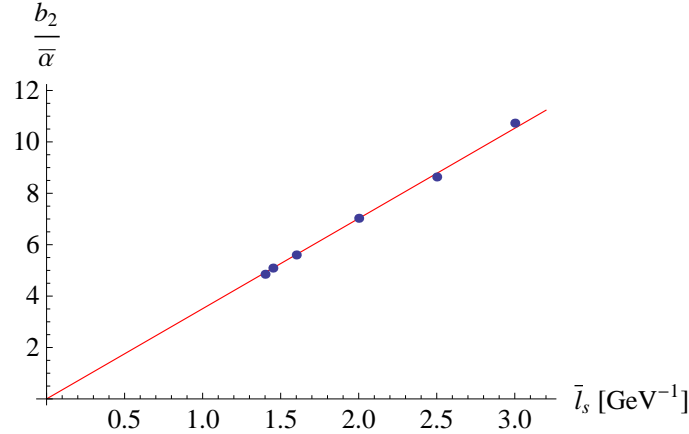}
  \end{center}
\caption{Functional dependence of $b_2/\bar\alpha$ versus $\bar l_s$
constrained by the condition that they give a physical value for the
string tension $\sigma = (0.425 \,{\rm GeV})^2$. We consider in this plot the region of the parameters set close to $b_2 = 2.3$. The line corresponds to Eq.~(\ref{eq:ccIR1}).}
\label{fig:plotb2a0}
\end{figure}

The parameters $\bar\alpha$, $b_2$ and $\bar l_s$ must be chosen to
  reproduce the physical value of the string tension $\sigma = (0.425
  \, {\rm GeV})^2$. From a numerical computation of the heavy
  $Q\bar{Q}$-potential in the regime $\rho\sim 5\,\textrm{GeV}^{-1}$,
  we find that these three parameters are constrained according to the relation
\begin{equation}
\frac{b_2}{\bar\alpha} = 3.51 \textrm{GeV} \cdot \bar l_s \,, \label{eq:ccIR1}
\end{equation}
as one sees in Fig.~\ref{fig:plotb2a0}.

The parameter $\bar l_s$ then follows from Eqs.~(\ref {eq:b2alpha})  and  (\ref{eq:ccIR1}), and we get:
\begin{eqnarray}
\bar l_s = 1.45 \,\textrm{GeV} ^{-1}\,. \label{eq:barls} 
\end{eqnarray}
This value is different from the string length $l_s =
2.62\,\textrm{GeV}^{-1}$ used in Section~\ref{sec:intro} with the
guessed metric.  Clearly a readjustment of the form of the metric may
also lead to a readjustment of the string length. At this point we
argue that the value of $\bar l_s$ is unambiguously fixed, and it is
not possible to accommodate a value of $\bar l_s$ equal to $l_s$
within our present analysis.

From our previous analysis and from Appendix B we see that a value
$1.5 < b_2 < 2.37$ and $ 0.29 < \bar\alpha < 0.47$ satisfying
Eq.~(\ref{eq:b2alpha}) ensures that the theory is confining and the
running coupling is well reproduced. Even when these intervals are
very narrow, one can desire to get concrete values for $b_2$ and
$\bar\alpha$. To this end we study the lowest $0^{++}$ and $2^{++}$
glueballs.  In the presence of both a gravity field and a dilaton
field a careful separation of the scalar degrees of freedom has to be
made~\cite{Kofman:2004tk,Kiritsis:2006ua}. We take from the second
reference~\cite{Kiritsis:2006ua} the corresponding effective
Schr\"odinger potential, which is given by

\begin{equation}
V_i^{\textrm{Schr.}}(z) = (B_i^\prime(z))^2 + B_i^{\prime\prime}(z)  \,, \qquad i=0,2 \,,
\end{equation}
where the functions $B_0(z)$ and $B_2(z)$ differ for the $0^{++}$ and  $2^{++}$ glueballs

\begin{eqnarray}
B_0(z) &=& \frac{3}{2} A(z)+\frac{1}{2}\log[ X^2(z)] \,,\\
B_2(z) &=& \frac{3}{2} A(z) \,.
\end{eqnarray}
$A(z)$ is the Einstein frame scale factor and $X[z] \equiv X[\alpha(z)]$ has been defined in Eq.~(\ref{exprX1}), being the dependence $\alpha(z)$ given by Eq.~(\ref{eq:dadz1}). Then we can solve the Schr\"odinger equation
\begin{equation}
\left[ -\frac{\partial^2}{\partial z^2} + V^{\textrm{Schr.}}(z)
\right] \psi_n(z) = m_n^2 \psi_n(z) \,.
\end{equation}
Best values for the glueballs $m_{0^{++}}=0.921 \,{\textrm GeV}$, $m_{2^{++}}=1.462\,\textrm{GeV}$ come out too low for  $b_2 = 2.3 $ and $ \bar \alpha = 0.45$, where $b_2$ and $\bar\alpha$ are constrained themselves according to Eq.~(\ref{eq:b2alpha}). These values for the parameters, which are close to the limit of good infrared singularity, cf. Appendix B, give optimum results for the glueball spectrum. The same happens in the fit of the running coupling in Sec.~\ref{modification}. This justifies the choice
\begin{eqnarray}
b_2 = 2.3 \,, \label{eq:b2} \\
\bar\alpha = 0.45 \label{eq:alphabar11} \,.
\end{eqnarray}
It is interesting to note that the glueball spectrum probes larges values of $\alpha$ in the dilaton potential than what the heavy $Q\bar{Q}$ potential does. For instance, at the energy of the ground state for $0^{++}$, the potential is tested up to values of $\alpha=11.1$, and for $2^{++}$ up to $\alpha=16.4$.

\begin{figure}[!ht]
  \begin{center}
     \epsfxsize 10cm \epsffile{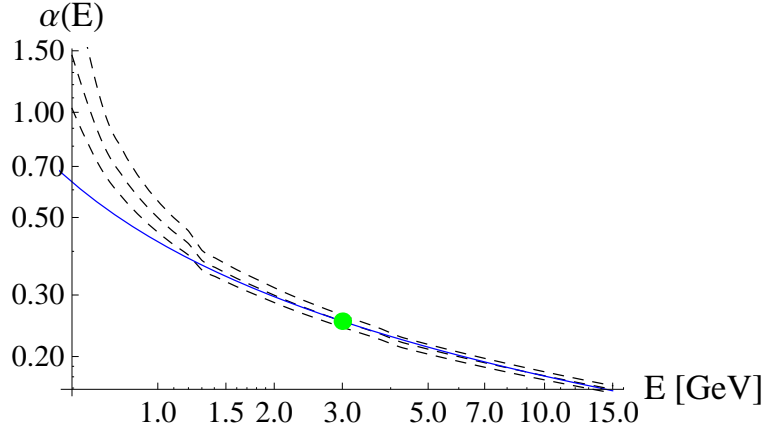}
  \end{center}
  \caption{\small The running coupling as a function of energy
    scale. We show as a full (blue) line our result stemming from the
    parametrization of Eq.~(\ref{newbeta}), with the values of the
    parameters given by Eqs.~(\ref{eq:b2}) and (\ref{eq:alphabar11}). The
    interpolation of the PDG data is displayed as dashed (black) lines,
    with its corresponding error. The point shows the input condition,
    Eq.~(\ref{eq:cond1alpha}).}
\label{Alpha-2}
\end{figure}

The resulting running coupling follows from the $\beta$-function of
Eq.~(\ref{newbeta}) by just using a single input
value~(\ref{eq:cond1alpha}). See the discussion after Eq.~(\ref{relation}).  The behavior of the running coupling and
its comparison to data from PDG is shown in Fig.~\ref{Alpha-2}. The
PDG values for the strong coupling are obtained via the PDG web
tool~\cite{PDG}. To check the numerical consistency, we compare in
Tab.~\ref{PDG_table} the values of the running coupling at several
energy scales before and after the modification of the dilaton
potential, with the corresponding values from PDG.  The $AdS/QCD$
model can match perturbative QCD-calculations with a very good
accuracy of about 1$\%$. This is important if one wants to connect a
perturbative Monte Carlo cascade with non-perturbative QCD physics in
parton fragmentation.

\begin{table}[!ht]
\begin{center}
\begin{tabular}{|c||c|c|c||c||c|}
\hline
Energy Scale & $\alpha_s $  & $+$ & $-$ & $\alpha $
& $\alpha $\\
$\mathrm{[GeV]}$ & (PDG) & {} & {} & Model 1: $h(z)$ & Model 2: $\bar h(z)$\\
\hline
$12$ & $0.16907$ & $0.00436$ & $0.00426$ & $0.16179$ & $0.16809$\\
\hline
$8.2$ & $0.18463$ & $0.00525$ & $0.00510$ & $0.18431$ & $0.18492$\\
\hline 
$5$ & $0.20994$ & $0.00688$ & $0.00662$ & $0.21371$ & $0.21279$\\
\hline
$2.6$ & $0.26708$ & $0.01178$ & $0.01106$ & $0.26809$ & $0.26640$\\
\hline
$2$ & $0.29942$ & $0.01522$ & $0.01408$ & $0.31023$ & $0.29668$\\
\hline
$1$ & $0.4996$ & $0.05799$ & $0.04598$ & $0.82393$ & $0.42618$\\
\hline
\end{tabular}
\end{center}
\caption{Values of the running coupling at different energy scales compared with PDG data~\cite{PDG}.}
\label{PDG_table}
\end{table}

The values of the running coupling provided by the old warp factor $h(z)$
of Eq.~(\ref{eq:hzPirner}) are doing quite well in comparison with the
ones given by PDG within the errors, but the values obtained from the
new dilaton potential~(\ref{analyticDilatonPotential}) are closer to
the experimental data.

Then only one parameter $V_0$ or $\bar\ell=\frac {6}{ \sqrt{-3 V_0}}$ remains to be fixed.
Focusing on the UV regime, we can perform an analytical study of the
short distance regime by expanding Eqs.~(\ref{finalrho}) and
(\ref{finalVqq}) in powers of $\alpha_0$. The details of the
computation are provided in Appendix~C. The result in NNLO is
\begin{equation}
V_{Q\bar{Q}}(\rho) = -\frac{2 \bar\ell^2}{\pi \bar l_s^2}
\frac{\alpha_0^{4/3}(\rho)}{\rho} 
\Big\{0.359 + 0.533 b_0 \alpha_0(\rho) + (1.347 b_0^2 + 0.692 b_1)
\alpha_0^2(\rho) + {\cal O}(\alpha_0^3)
\Big\} \,. \label{eq:Valphatext1}
\end{equation}
By reversing Eq.~(\ref{finalrho}), cf. Eq.~(\ref{eq:rhoexpand1}) in Appendix~C,
we get the following functional form for
$\alpha_0$ as a function of the separation $\rho$
\begin{eqnarray}
\alpha_0(\rho) &=& \frac{1}{b_0 \log\left(\frac{1.32\bar\ell}{\rho}\right) +\frac{b_1}{b_0} \log \left( b_0 \log \left( \frac{1.32\bar\ell}{\rho} \right) \right)}  
  - \frac{\left( 0.079 b_0^2  +\frac{b_1^2}{b_0^2}\right)}{\left(b_0 \log\left(\frac{1.32\bar\ell}{\rho}\right) +\frac{b_1}{b_0} \log \left( b_0 \log\left( \frac{1.32\bar\ell}{\rho} \right) \right) \right)^3} \nonumber \\
  && + {\cal O}\left( \log^{-4} \left(\frac{1.32\bar\ell}{\rho}\right) \right) \,. \label{eq:alpharhotext1}
\end{eqnarray}

The heavy quark potential can be directly compared with perturbation
theory (PT). $V_{Q\bar{Q}}$ is computed in PT as an
expansion in powers of the QCD running coupling~\cite{Brambilla:2004jw}. It has
the form 
\begin{equation}
  V_{\textrm{PT}}(\rho) = -\frac{N_c^2-1}{2 N_c}\cdot \frac{\alpha_V(\rho)}{\rho},
\label{PotentialForm-1}
\end{equation}
where up to the third order
\begin{eqnarray}
  \alpha_V &=& \alpha_\text{\tiny PT} \bigg\{1 + (a_1 +
4\pi\gamma_E b_0)\frac{\alpha_\text{\tiny PT}}{4\pi} \nonumber \\
&&+\left[4\pi\gamma_E(2a_1 b_0 + 4\pi b_1) + 4\pi^2\left(\frac{\pi^2}{3} -
4\gamma_E^2\right)b_0^2 + a_2\right]
\frac{\alpha_\text{\tiny PT}^2}{16\pi^2} \bigg\} \,. \label{eq:alphav}
\end{eqnarray}
In this expression $b_0$ and $b_1$ were defined in
Eq.~(\ref{QCDbeta}), $\alpha_\text{\tiny PT}$ is the perturbative QCD
running coupling, and the coefficients $a_1$ and $a_2$ were calculated
by Fischler \cite{Fischler:1977yf} and by Peter \cite{Peter:1996ig}
and Schr\"oder \cite{Schroder:1998vy}, respectively. We use the convention of
Ref.~\cite{Brambilla:2004jw} and Ref.~\cite{Schroder:1998vy}:
\begin{equation}
  a_1=\frac{31}{9}N_c-\frac{10}{9}N_f,
\end{equation}
\begin{eqnarray}
  a_2 &=& \left(\frac{4343}{162}+4\pi^2-\frac{\pi^4}{4}+\frac{22}{3}\zeta(3)
\right)N_c^2 - \left(\frac{5081}{324} + \frac{16}{3}\zeta(3) \right)
N_c N_f  \nonumber \\
&& + \left( \frac{55}{12} - 4\zeta(3) \right) \frac{N_f}{N_c} +
\frac{100}{81}N_f^2.
\end{eqnarray}
Practically, we use the two-loop running coupling constant, which
has the following form
\begin{equation}
  \alpha_\text{\tiny PT} (\rho, d) = \frac{1}{b_0\cdot
\mathrm{log}\left(\frac{d}{\rho\cdot \Lambda}\right) +
\frac{b_1}{b_0}\cdot \mathrm{log}\left(2\mathrm{log}
\left(\frac{d}{\rho\cdot \Lambda}\right)\right)}, 
\label{eq:alphaPT}
\end{equation}
where $d$ is an undetermined parameter which relates the scale~$\mu$ of the
running coupling with the distance $\rho$, i.e. $\mu = d/\rho$.

The expansion (\ref{eq:Valphatext1}) is similar to that of PT,
Eqs.~(\ref{PotentialForm-1})-(\ref{eq:alphav}), except that there is
an extra power $\alpha_0^{1/3}$ at every order. This is a common
prediction of all the renormalization group revised models constructed
by the general procedure of Kiritsis et al., cf. Refs.~\cite{Gursoy:2007cb,Gursoy:2007er}. 
At first sight this difference is
a matter of concern.  In order to fix the unknown parameter  $V_0$ of the
dilaton potential, we proceed in the following way: We find
that Eq.~(\ref{eq:Valphatext1}) depends on the factor $\bar\ell^2/\bar l_s^2 =
-12/(V_0 \bar l_s^2)$, and a numerical comparison between the leading orders in
Eqs.~(\ref{eq:Valphatext1}) and (\ref{PotentialForm-1}) in the regime $0.06
\,\textrm{GeV}^{-1} < \rho < 0.20 \,\textrm{GeV}^{-1}$ gives us the value $-V_0 \bar l_s^2 = 1.31$, which is confirmed in a big range of $\bar l_s$. 
Using the determined value of $\bar l_s$ we get:
\begin{equation}
V_0 = -0.623\,\textrm{GeV}^2 \,.
\end{equation}

A convenient choice for the parameter $d$ follows from a direct comparison between the argument inside the logarithms in
Eqs.~(\ref{eq:alpharhotext1}) and (\ref{eq:alphaPT}),
\begin{equation}
d = 1.32 \bar\ell \Lambda = 1.53 \,.
\end{equation}
We choose the value of $\Lambda$ given in
Eq.~(\ref{eq:Lambda}). Another possibility would be to use a $\Lambda$
which follows from QCD studies for the running coupling.  In either
case the value of $d$ is chosen in such a way that it compensates a
change in $\Lambda$.

\begin{figure}[!ht]
  \begin{center}
    \epsfxsize 9cm \epsffile{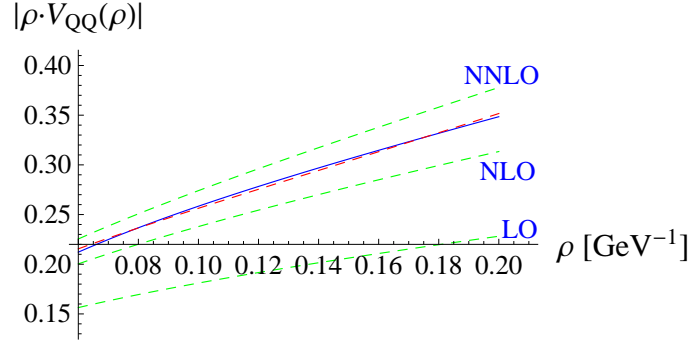}
  \end{center}
\caption{The heavy quark-antiquark potential as a function of the
distance $\rho$ for small $\rho$. The result stemming
from the dilaton potential of Eq.~(\ref{analyticDilatonPotential}) is
shown as a full (blue) line. It follows from a numerical
computation of Eqs.~(\ref{finalrho}) and (\ref{finalVqq}). The
perturbative computation, Eqs.~(\ref{PotentialForm-1}) and
(\ref{eq:alphav}), is displayed as a dashed (red) line. The short
distance expansion, Eq.~(\ref{eq:Valphatext1}), is displayed up to
leading order (LO), next-to-LO and next-to-next-to-LO, as dashed
(green) lines from bottom to top, respectively. We consider in this
plot $\bar l_s = 1.45\,\textrm{GeV}^{-1}$ and $V_0 = -0.623 \,\textrm{GeV}^2$.}
\label{fitVqq-3}
\end{figure}

We compare in Fig.~\ref{fitVqq-3} the numerical
result for the heavy quark-antiquark potential from
Eq.~(\ref{finalVqq}), with the perturbative result of QCD given by
Eqs.~(\ref{PotentialForm-1}) and (\ref{eq:alphav}), in the small
distance regime $0.06 \,\textrm{GeV}^{-1} < \rho < 0.20
\,\textrm{GeV}^{-1}$. The upper bound in $\rho$ is motivated by the fact that
our result for the running coupling fits well the experimental data in
the regime $E > 4-5 $~GeV. We also show in dashed lines the short
distance expansion of Eq.~(\ref{eq:Valphatext1}) up to leading order
${\cal O}(\alpha_0^{4/3})$, next-to-leading order ${\cal
O}(\alpha_0^{7/3})$ and next-to-next-to-leading order ${\cal
O}(\alpha_0^{10/3})$. We see that the comparison with the numerical
computation of $V_{Q\bar{Q}}$ and perturbation theory is quite
accurate, although it seems to be that the series
(\ref{eq:Valphatext1}) is slowly convergent. It is important to note
that the full string potential is rather close to the perturbative
potential in spite of the expansion containing different powers in the
Yang-Mills coupling.

\begin{figure}[!ht]
  \begin{center}
    \epsfxsize 10cm \epsffile{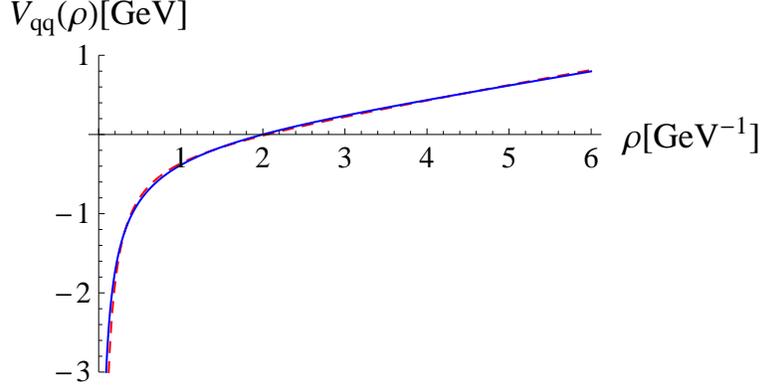}
  \end{center}
\caption{The heavy quark-antiquark potential as a function of
$\rho$. Our result stemming from the dilaton potential of
Eq.~(\ref{analyticDilatonPotential}) is shown as a full (blue)
line. It follows from a numerical computation of
Eqs.~(\ref{finalrho}) and (\ref{finalVqq}).  The dashed (red) line
corresponds to the Cornell potential, Eq.~(\ref{CornellForm}).}
\label{Vqq}
\end{figure}

A full numerical computation gives the heavy
$Q\bar{Q}$-potential from Eqs.~(\ref{finalrho}) and
(\ref{finalVqq}). The result is shown in Fig.~\ref{Vqq} .
One can try to fit our result with the Cornell form of the potential:
\begin{equation}
  V^{\textrm{Cornell}}_{Q\bar{Q}}(\rho)=-\frac{a}{\rho}+\sigma\cdot\rho+C  \,,
\label{CornellForm}
\end{equation}
and obtains the values
\begin{displaymath}
  a = 0.42,\quad \sigma = (0.415\,\mathrm{GeV})^2,\quad C =
-0.14\,\mathrm{GeV},
\end{displaymath}
which are rather close to the accepted values~\cite{Eichten:1978tg,Eichten:1979ms}, and also to the values obtained in Ref.~\cite{Pirner:2009gr}. In the numerical computation of Eq.~(\ref{finalVqq}) we have conveniently normalized the result by adding a constant $C$ in order that the $Q\bar{Q}$-potential vanishes close to $\rho = 2\, \textrm{GeV}^{-1}$.

\section{Dilaton Potential and New Warp Factor}
\label{sec:dilaton-runningcoupling}

With the three parameters $V_0, \bar{\alpha}$ and $ b_2$ determined in the
previous section we can obtain the dilaton potential Eq.~(\ref{analyticDilatonPotential}), the running coupling and the modified warp factor $\bar h(z)$ of Eq.~(\ref{relation}). 

The parameters governing the UV-asymptotic
behavior of the dilaton potential are $b_0=\frac{25}{6\pi}$
and $b_1=\frac{77}{12\pi^2}$, the known coefficients of the
perturbative $\beta$-function, in addition to $V_0$, as can be seen in Eq.~(\ref{eq:Vdilaton}). So, the potential is consistent with the QCD $\beta$-function at two-loop level.

\begin{figure}[!ht]
  \begin{center}
     \epsfxsize 10cm \epsffile{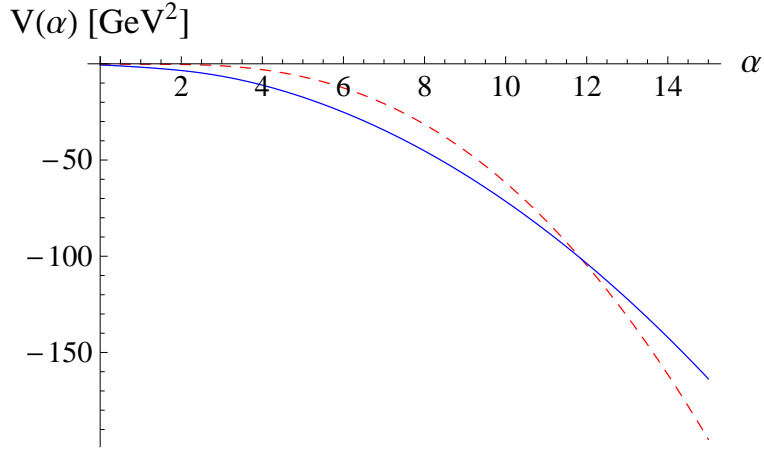}
  \end{center}
  \caption{\small The dilaton potential as a function of the running coupling $\alpha$. The result corresponding to the parametrization of Eq.~(\ref{eq:Vdilaton}) is displayed as a continuous (blue) line. We show for comparison as a dashed (red) line the potential stemming from the warp factor $h(z)$ of Eq.~(\ref{eq:hzPirner}), computed in Sec.~\ref{sec:soldilaton}.}
\label{newPotential}
\end{figure}

The dilaton potential is plotted in Fig.~\ref{newPotential} and compared
with the one obtained from the warp factor $h(z)$ of
Eq.~(\ref{eq:hzPirner}). The comparison shows that the modified dilaton potential becomes slightly flatter. This flattening is in line with the result given by Ref.~\cite{Gursoy:2009jd}, although the parameters provided in this reference together with our value of $\bar l_s$ produce a potential of the order of $10^4\, \textrm{GeV}^2$ in the regime of physical interest, $\alpha \approx 0.3$, in contrast to the value $\sim 1 \, \textrm{GeV}^2$ given by our potential.

%%%%%%%%%%%%%%%%%%%%%%%%%%%%%%%%%%%%%%%%%%%%%%%%%%%%%%%%%%%%%%%%%%%%%%%%%%

It is interesting to compute the warp factor $\bar
h(z)$ that follows from the new dilaton potential of
Eq.~(\ref{analyticDilatonPotential}). This way we can close the circle
of investigation in the present paper. The mathematical procedure is
as follows: From Eqs.~(\ref{eq:dadz1}) we get a
first order differential equation

\begin{equation}
\frac{d\alpha}{dz} = \frac{\sqrt{-3 V_0}}{6} e^{A(\alpha)-D(\alpha)} \,, \label{eq:dadz}
\end{equation}
where $A(\alpha)$ and $D(\alpha)$ are given by Eqs.~(\ref{eq:Aalpha})
  and (\ref{eq:eDDD}) respectively. This equation can be solved by
  imposing the boundary condition $\alpha(z \to 0) \to 0$, due to the
  asymptotic freedom.  The solution is delicate, because there are
  problems to find a numerical solution of Eq.~(\ref{eq:dadz}) near
  the boundary $z=0$.  At this point the l.h.s.  of
  Eq.~(\ref{eq:dadz}) is divergent. To overcome this difficulty, we
  proceed in three steps. First, we consider the lowest order
  perturbative expansion of the $\beta$-function and rewrite
  Eq.~(\ref{eq:dadz}) in the deep UV,

\begin{equation}
\frac{d\alpha_{\textrm{\tiny UV}}}{dz} = b_0 \frac{\sqrt{-3 V_0}}{6} \, \alpha_\textrm{\tiny UV}^2
\, e^{\frac{1}{b_0 \alpha_\textrm{\tiny UV}}} \,. \label{eq:dadzuv}
\end{equation}
The solution of this equation is:

\begin{equation}
\alpha_{\textrm{\tiny UV}} (z) = -\frac{1}{b_0 \log \left(\bar\Lambda z \right)} \,,
\end{equation}
with 

\begin{equation}
\bar\Lambda = \frac{\sqrt{-3 V_0}}{6} = 237 \,\textrm{MeV}\,.
\end{equation}
Note that the value of $\bar\Lambda$ is rather close to
the one determined using $h$, cf. Eq.~(\ref{eq:Lambda}). 
This small discrepancy could be improved when higher orders in the
perturbative expansion are considered, as it has been observed in the
analy\-tical computation of the $Q\bar{Q}$-potential, see Appendix~B,
cf. Eqs.~(\ref{eq:apprhosmall}) and (\ref{eq:Appc}).

Choosing $z_0=1.2 \cdot 10^{-4} \, \textrm{GeV}^{-1}$ we obtain
$\alpha(z_0)= 0.0718$
as initial value  in the deep UV
to find the numerical solution of
Eq.~(\ref{eq:dadz}). In the second step 
we consider three orders in the UV expansion of the
$\beta$-function, Eq.~(\ref{newbeta}), and $A(\alpha)$,
and solve Eq.~(\ref{eq:dadz}) from $z_0=1.2\cdot 10^{-4}\, \textrm{GeV}^{-1}$ to $z_1= 0.125\,\textrm{GeV}^{-1}$ with

\begin{equation}
A(\alpha) = C_A + \frac{1}{b_0 \alpha} + \frac{b_1}{b_0^2} \log \alpha
+ \frac{1}{6 b_0^3 \bar\alpha^3} \left(b_0 b_2 - 3 b_0^2 \bar\alpha -
6 b_0 b_1 \bar\alpha^2 - 6 b_1^2 \bar\alpha^3\right) \alpha + {\cal
O}(\alpha^2) \,, \label{eq:Aexpansion}
\end{equation}
where $C_A= -0.1057$ is a constant to make $A(\alpha)$ consistent with
the input condition~(\ref{eq:cond1alpha}). Finally from $z_1$ to
infinity we use the input functions to solve the equation fully
numerically. We have checked the stability of the solution changing
$z_0$ and $z_1$, and considering higher orders in the
expansion, Eq.~(\ref{eq:Aexpansion}).  Once we know $\alpha(z)$, the
corresponding warp factor is easily computed, and it reads

\begin{equation}
\bar h(z) = e^{2 A(\alpha(z))} (\alpha(z))^{\frac{4}{3}} (\Lambda z)^2 \,.
\end{equation}

\begin{figure}[!ht]
  \begin{center}
     \epsfxsize 10cm \epsffile{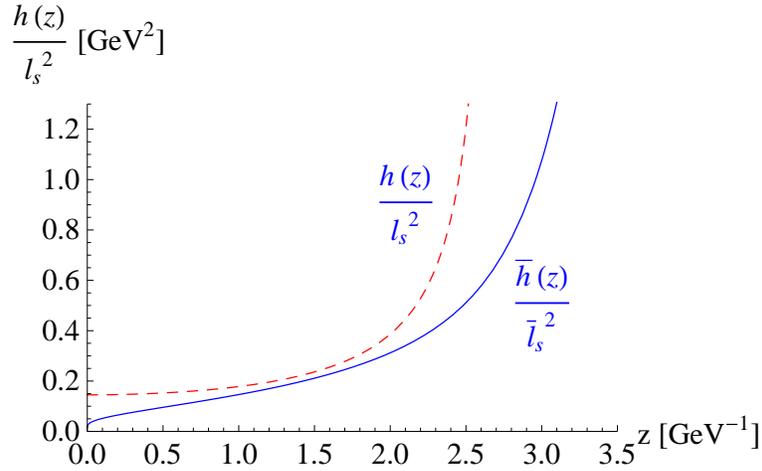}
  \end{center}
  \caption{\small The warp factor divided by $l_s^2$ as a function of $z$. The factor of Eq.~(\ref{eq:hzPirner}) proposed in Ref.~\cite{Pirner:2009gr} is shown as a dashed (red) line. The full line corresponds to the warp factor that follows from the dilaton potential of Eq.~(\ref{analyticDilatonPotential}) using the parameters determined in Sec.~\ref{calculation_after_modification}.}
\label{fig:hz}
\end{figure}

The strength in the Nambu-Goto action, Eq.~(\ref{eq:NG1}), is
determined by the factor $h(z)/l_s^2$, and so it is more relevant to
consider this quantity when comparing different models. The result
$\bar h(z)/\bar l_s^2$ is shown in Fig.~\ref{fig:hz} and compared to
the warp factor of Eq.~(\ref{eq:hzPirner}) that was first proposed in
Ref.~\cite{Pirner:2009gr}. $l_s^2$ is approximately a factor $3$
larger than $\bar l_s^2$, and this is reflected also in the values of
$h(z)$ compared to $\bar h(z)$. The numerical agreement is rather good
up to $z \simeq 2 \,\textrm{GeV}^{-1}$, in spite of the fact that
$\bar h(z)$ vanishes in the UV. $\bar h(z)$ has a singularity at $z_{\textrm {\tiny IR}} = 3.65 \, \textrm{GeV}^{-1}$, which occurs at slightly larger values than in $h(z)$, for which $z_{\textrm {\tiny IR}} = \sqrt{1-\epsilon}/\Lambda = 2.73 \, \textrm{GeV}^{-1}$.

In the deep UV, i.e. at leading order in the
perturbative expansion of the $\beta$-function, the analytical
solution of Eq.~(\ref{eq:dadz}) has a simple form:

\begin{eqnarray}
\bar h_{\textrm{\tiny UV}} (\alpha) &=& \left( \frac{\Lambda}{\bar\Lambda}
\right)^2 \alpha_{\textrm{\tiny UV}}^{\frac{4}{3}} \\
                              &=&  \left( \frac{\Lambda}{\bar\Lambda}
\right)^2 \frac{1}{\left( - b_0 \log \left( \bar \Lambda z \right)\right)^{\frac{4}{3}}} \,.
\end{eqnarray}
This functional form  of the calculated warp factor is very similar
to the guessed warped factor besides the power $4/3$.

\section{Discussion and final remarks}
\label{sec:discussion} 
The analogy of the bulk coordinate $z$ with the inverse energy
resolution has triggered the guessed warp factor in
Refs.~\cite{Pirner:2009gr,Nian:2009mw}, which was based on a naive
equivalence with the running coupling of QCD. A careful analysis of
the resulting dilaton potential gives the evolution of the dilaton
field in the bulk and consequently the running of the QCD gauge
coupling. The infrared physics of the model of
Refs.~\cite{Pirner:2009gr,Nian:2009mw} was satisfactory to fit the
string tension, but it fails to give a good UV-behavior for the
$\beta$-function. Because of the second order Einstein equations and
the correlated behavior of the dilaton in the infrared and
ultraviolet, which is not constrained by the correct QCD
$\beta$-function, one needs two boundary conditions to interpolate the
gauge coupling between the charmonium and bottonium masses. This
feature weakens the idea of holography for the old ansatz, which
determines the field theoretic behavior of our 4-dimensional world
from the physics of gravity in 5-dimensional anti-de Sitter space.

In order to have the Yang-Mills theory as a `hologram' of the physics
happening in five dimensions we assumed a new ansatz that improves the
UV-behavior by using the QCD $\beta$-function as a
constraint. Thereby we found a dilaton potential which is consistent
with QCD in the UV region. The resulting short distance heavy quark
potential $r\cdot V_{Q\bar{Q}}(r)$ has a similar shape as the 3-loop
expression derived by Brambilla et al. \cite{Brambilla:2004jw}. The
numerical comparison of the QCD and string heavy $Q\bar{Q}$-potential
is rather good in spite of the fact that the leading term in string
theory proportional to $\alpha^{4/3}$ deviates from the QCD-potential
proportional to $\alpha$. By calculating the NNLO-expansion, we show
that the expansion in $\alpha$ is slowly converging.  With a new
string length~$\bar l_s$ in the Nambu-Goto action we can also match
the long range $Q\bar{Q}$-interaction. We have closed the circle of
investigation by computing the warp factor corresponding to the new
ansatz, and found that the scaled warp factor is similar to that of
Refs.~\cite{Pirner:2009gr,Nian:2009mw} in the region of interest $ 0.5
\,\textrm{GeV}^{-1} \le z \le 2 \, \textrm{GeV}^{-1}$.

In the procedure we have proposed, we fix three parameters $(b_2,{\bar\alpha,V_0})$ corresponding to the dilaton potential and one parameter $\bar l_s$ equal to the string length by using three constraints, namely a good behavior of the $Q\bar{Q}$-potential in the IR and in the UV, and a good behavior of the running coupling in the regime $0.6 \,\textrm{GeV } \le E \le 15 \,\textrm{GeV}$. New parameters may have to be included in order to describe the glueball spectrum or further observables with precision. One important point of this analysis is the general criterion for confinement of Ref.~\cite{Gursoy:2007er} and the requirement that the infrared singularity being repulsive to physical modes, which help us to set a narrow window in our parameter set, in particular for the parameters controlling the infrared behavior of the theory $(b_2, \bar\alpha)$. 

Recently the question ``How Well Does AdS/QCD Describe QCD?''
\cite{Erlich:2009me} has been asked, and depending on the feature the
answer varied - the accuracy was estimated between $10\%$ and
$25\%$. In our case of pure gluon dynamics, we have shown that the
accuracy is much better, and therefore we look optimistically towards
further tests of the action in finite temperature calculations
\cite{Nian:2010}.

\vspace{1cm}

{\bf Acknowledgments:}

E.M. would like to thank the Humboldt Foundation for their stipend. This work was also supported in part by the ExtreMe Matter Institute EMMI in the framework of the Helmholtz Alliance Program of the Helmholtz Association.

\newpage

\appendix{\textsf{\large Appendix A: Infrared and Ultraviolet Properties of the Gravity Dual Theory}}
\label{sec:apA}

In this appendix we study the infrared (IR) and the ultraviolet (UV) properties of the 5-dim Nambu-Goto theory with the ``guessed'' metric $h(z)/(\Lambda z)^2$, cf. Eq.~(\ref{eq:hzPirner}), which we develop in Secs.~\ref{gravity} and \ref{sec:soldilaton}. We apply technical details of Refs.~\cite{Gursoy:2007cb,Gursoy:2007er}, which are shortly reviewed in Sec.~\ref{modification}. The warp factor $h(z)$ has a singularity at $z_{\textrm {\tiny IR}} = \sqrt{1-\epsilon}/\Lambda$, so the bulk coordinate $z$ is restricted to $z<z_{\textrm {\tiny IR}}$. The IR expansion of $h(z)$ is
\begin{equation}
h(\xi) = \frac{\log(\frac{1}{\epsilon})}{2 \sqrt{1-\epsilon}}\frac{1}{\Lambda\xi} - \frac{(1-2\epsilon) \log(\frac{1}{\epsilon})}{4 (1-\epsilon)} + {\cal O}(\xi) \,, \qquad \xi \to 0 \,, \label{eq:hIR}
\end{equation}
with
\begin{equation}
\xi = z_{\textrm {\tiny IR}} - z \,.
\end{equation}
Using Eqs.~(\ref{relation1}) and (\ref{EEOM2}) we get a second order differential equation for $\phi(\xi)$, which writes
\begin{equation}
\phi^{\prime\prime}(\xi) + \left( \frac{1}{\xi} - \frac{3+2\epsilon}{2\sqrt{1-\epsilon}}\Lambda + {\cal O}(\xi) \right) \phi^\prime(\xi) -\frac{3}{8} \left(\frac{1}{\xi^2} + \frac{3+2\epsilon}{\sqrt{1-\epsilon}}\frac{\Lambda}{\xi} + {\cal O}(\xi^0)\right) = 0 \,. 
\end{equation}
This equation can be solved for several orders in the IR expansion of $h(\xi)$. The theory of differential equations gives the general solution by adding to the special solution of the inhomogeneous equation the full set of homogeneous solutions. The result is
\begin{equation}
\phi(\xi) = \frac{3}{16}(\log\xi)^2 + c_1 \log\xi + c_2 + {\cal O}(\xi\log\xi) \,. \label{eq:phiir1}
\end{equation}
The parameters $c_1$ and $c_2$ are two unknown constants corresponding to the homogeneous solutions which have to be fixed by two conditions. Obviously the three terms that we show explicitly in Eq.~(\ref{eq:phiir1}) correspond to the lowest orders of the homogeneous and inhomogeneous solutions. At this point it is preferable to write the solution in this way
\begin{equation}
\phi(\xi) = \frac{3}{16}\left(\log\frac{\xi}{\omega_{\textrm {\tiny IR}}}\right)^2 + \kappa_{\textrm{\tiny IR}} + {\cal O}(\xi\log\xi) \,, \label{eq:phiir2}
\end{equation} 
where the constants $\omega_{\textrm {\tiny IR}}$ and $\kappa_{\textrm{\tiny IR}}$ are related to $c_1$ and $c_2$ as
\begin{equation}
\omega_{\textrm {\tiny IR}} = e^{-\frac{8}{3} c_1} \,, \qquad \kappa_{\textrm{\tiny IR}} = -\frac{4}{3} c_1^2 + c_2 \,.
\end{equation}
Setting $\omega_{\textrm {\tiny IR}}$ corresponds to setting the scale. We can solve Eq.~(\ref{EEOM2}) numerically for the full range of $z<z_{\textrm {\tiny IR}}$ using the IR behavior of Eq.~(\ref{eq:phiir2}) as a boundary condition. In the concrete calculation we choose $\phi(\xi_1)$ and $\phi^\prime(\xi_1)$ with $\xi_1$ very small. Doing that, we have checked that the result of Fig.~\ref{Dilatonprofile} is exactly reproduced for
\begin{equation}
\omega_{\textrm {\tiny IR}} = 4.55 \;{\rm GeV}^{-1} \,, \qquad  \kappa_{\textrm{\tiny IR}} = -0.758   \,. \label{eq:xi0kappa}
\end{equation}
These numbers are stable in the deep infrared near the singularity, $\xi_1 \sim 10^{-7} - 10^{-4} \, {\rm GeV}^{-1}$. So, this choice of the constants $\omega_{\textrm {\tiny IR}}$ and $\kappa_{\textrm{\tiny IR}}$ is equivalent to the boundary conditions of~Eq.~(\ref{runningcouplingvalues}).

The IR behavior of $A(\xi)$ can be obtained from Eqs.~(\ref{relation1}), (\ref{eq:hIR}) and (\ref{eq:phiir2}), and it reads
\begin{equation}
A(\xi) = -\frac{1}{8} \left( \log\frac{\xi}{\omega_{\textrm {\tiny IR}}} \right)^2 -\frac{1}{2}\left(\log\frac{\xi}{\omega_{\textrm {\tiny IR}}}\right) -\frac{2}{3}\kappa_{\textrm{\tiny IR}} + \frac{1}{2} \log\left(\frac{\log\frac{1}{\epsilon}}{2(1-\epsilon)^{\frac{3}{2}} \omega_{\textrm {\tiny IR}} \Lambda} \right) + {\cal O}(\xi \log\xi) \,, \quad \xi \to 0 \,.\label{eq:Air}
\end{equation}
Once that we know $A(\xi)$, the IR asymptotics of the superpotential $W$ can be computed using Eq.~(\ref{defW}) and taking into account the domain wall coordinates relation $e^A dz = du$. It~reads, 
\begin{equation}
W(\alpha) = W_\infty \,\alpha^{\frac{2}{3}} \left(\log\alpha\right)^\frac{1}{2} e^{\frac{2}{\sqrt{3}} \sqrt{\log\alpha} } \left(1 -\frac{3+2\kappa_{\textrm{\tiny IR}}}{2\sqrt{3}}\frac{1}{(\log\alpha)^{\frac{1}{2}}}  +  {\cal O}\left(\left(\log\alpha\right)^{-1}\right) \right) \,, \quad \alpha\to\infty \,, \label{eq:Wir}
\end{equation}
where the constant $W_\infty$ is
\begin{equation}
W_\infty = \frac{3}{4} \sqrt{\frac{6(1-\epsilon)^{\frac{3}{2}}}{\log\left(\frac{1}{\epsilon}\right)} \frac{\Lambda}{\omega_{\textrm {\tiny IR}}}} \,.
\end{equation}
The dependence in $\kappa_{\textrm{\tiny IR}}$ appears at ${\cal O}((\log\alpha)^{-\frac{1}{2}})$ in the bracket of Eq.~(\ref{eq:Wir}). The fact that the superpotential $W(\alpha)$ grows faster than $\alpha^{\frac{2}{3}}$ in the infrared ensures that the theory is confining and also that there is a mass gap in the spectrum~\cite{Gursoy:2007er,Gursoy:2008za}. Note that this is true independently of the values of $\omega_{\textrm {\tiny IR}}$ and $\kappa_{\textrm{\tiny IR}}$.
 
The dilaton potential can be obtained from $W$ using the third relation of Eq.~(\ref{newEinstein}), or equivalently from $A(z)$ and Eq.~(\ref{EEOM1}). In this regime it behaves as
\begin{equation}
V(\alpha) = V_\infty \,\alpha^{\frac{4}{3}}(\log\alpha) \; e^{\frac{4}{\sqrt{3}}\sqrt{\log\alpha}} \left( 1 - \frac{2(2+\kappa_{\textrm{\tiny IR}})}{\sqrt{3}}\frac{1}{(\log\alpha)^{\frac{1}{2}}} + {\cal O}\left((\log\alpha)^{-1} \right) \right)  \,, \qquad \alpha\to\infty \,, \label{eq:Vir}
\end{equation}
where the constant $V_\infty$ is
\begin{equation}
V_\infty = -\frac{16}{9} W_\infty^2 \,.  \label{eq:VWinfty}
\end{equation}
The general form of the potential in the infrared that has been studied in Refs.~\cite{Gursoy:2007cb,Gursoy:2007er} is
\begin{equation}
V(\alpha) \sim \alpha^{2Q} (\log\alpha)^P \,, \qquad \alpha \to \infty \,. \label{eq:Vgursoy}
\end{equation}
The solution of Eq.~(\ref{eq:Vir}) corresponds to $Q=2/3$ and $P=1$ in this notation, in addition to an extra factor $e^{\frac{4}{\sqrt{3}}\sqrt{\log\alpha}}$ which is dominant with respect to $\log\alpha$, but subdominant with respect to $\alpha^{\frac{4}{3}}$. This subdominance, in addition to the fact that $Q < 2 \sqrt{2}/3$ ensures that the  IR singularity is of the good kind according to the criterion of G\"ursoy et al.~\cite{Gursoy:2007er}, which means that the singularity should be repulsive to physical fluctuations. To avoid any doubt about the extra factor, it is possible to prove that it can be removed in Eqs.~(\ref{eq:Wir}) and (\ref{eq:Vir}) by shifting the dilaton field 
\begin{equation}
\bar\phi^\frac{1}{2} = \phi^\frac{1}{2} + \frac{\sqrt{3}}{2} \,,
\end{equation}
which doesn't affect the IR assymptotics. The resulting expression for the assymptotics of the dilaton potential is:
\begin{equation}
V(\bar\alpha) \sim \bar\alpha^\frac{4}{3} \log\bar\alpha \,, \qquad \bar\alpha \to \infty \,.
\end{equation}

Note that the asymptotics of Eq.~(\ref{eq:Air}) has not been studied by the authors of Ref.~\cite{Gursoy:2007cb,Gursoy:2007er}, and so this case is not listed in Table~1 of Ref.~\cite{Gursoy:2007er}. One can add our result to this table. The IR behavior of $X(\alpha)$ can be computed from Eqs.~(\ref{defX}) and~(\ref{eq:Wir}):
\begin{equation}
X(\alpha) = -\frac{1}{2} -\frac{\sqrt{3}}{4} \frac{1}{(\log\alpha)^\frac{1}{2}}- \frac{3}{8} \frac{1}{\log\alpha}  - \frac{\sqrt{3}(3+2\kappa_{\textrm{\tiny IR}})}{16} \frac{1}{(\log\alpha)^{\frac{3}{2}}}   + \dots \,, \qquad \alpha\to\infty \,. \label{eq:Xiranalytic}
\end{equation}
The term $\propto 1/(\log\alpha)^{\frac{1}{2}}$ comes from the extra factor $e^{\frac{2}{\sqrt{3}} \sqrt{\log\alpha} }$ in $W(\alpha)$. In our case, the limit
\begin{equation}
\lim_{\alpha\to\infty} \left(  X(\alpha) + \frac{1}{2}\right) \log\alpha = K \,,
\end{equation} 
leads to $K=-\infty$. 

One important aspect of this analysis is that one has to fix two integration constants, $\omega_{\textrm {\tiny IR}}$ and $\kappa_{\textrm{\tiny IR}}$, using initial conditions. This contrasts with the analysis of Refs.~\cite{Gursoy:2007er,Gursoy:2008za}, where they show that just one initial condition is enough. One possibility studied in Refs~\cite{Gursoy:2007er,Gursoy:2008za} is fixing one of the parameters by requiring that the bulk singularity is not of the ``bad kind'', which means that the singularity should be repulsive to physical fluctuations, cf. Eq.~(E.28) of Ref.~\cite{Gursoy:2008za}. In our case the singularity of our solution at $z=z_{\text {\rm IR}}$ is of the good kind, independently of the values of $\omega_{\textrm {\tiny IR}}$ and $\kappa_{\textrm{\tiny IR}}$, cf. Eqs.~(\ref{eq:Wir}), (\ref{eq:Vir}), (\ref{eq:VWinfty}) and compare with Eqs.~(E.27) and (E.29) of Ref.~\cite{Gursoy:2008za}. To give a new perspective to this issue, we can study the UV behavior of the dilaton using $h(z)$ of Eq.~(\ref{eq:hzPirner}). Following the same procedure that we explained for the IR but considering an expansion at small $z$, one gets
\begin{equation}
\phi(z) = -\frac{\omega_{\textrm {\tiny UV}}}{z} + \kappa_{\textrm{\tiny UV}} + \frac{(\omega_{\textrm {\tiny UV}}  \Lambda)^2}{\epsilon \log\left(\frac{1}{\epsilon}\right)} \frac{z}{\omega_{\textrm {\tiny UV}}} + {\cal O}(z^2) \,, \qquad z\to 0 \,, \label{eq:phiuv}
\end{equation}
where $\omega_{\textrm {\tiny UV}}$ is a parameter setting the scale and $\kappa_{\textrm{\tiny UV}}$ is another parameter. They play the role of $\omega_{\textrm {\tiny IR}}$ and $\kappa_{\textrm{\tiny IR}}$ respectively in the UV. In fact these parameters are related, in the sense that setting $\kappa_{\textrm{\tiny UV}}$ in the UV then sets $\kappa_{\textrm{\tiny IR}}$ in the IR (the same for $\omega_{\textrm {\tiny IR}}$ and $\omega_{\textrm {\tiny UV}}$), cf. Sec.~\ref{sec:soldilaton}. From Eq.~(\ref{eq:phiuv}) and using the same procedure that we explain above, one gets
\begin{equation}
\beta(\alpha) = -\frac{3}{2}\alpha - \frac{9}{4} \frac{\alpha}{\log\alpha} -\frac{9(3+2\kappa_{\textrm{\tiny UV}})}{8}\frac{\alpha}{(\log\alpha)^2} +  {\cal O}\left(\frac{\alpha}{(\log\alpha)^3}\right) \,, \label{eq:betahPirner}
\end{equation}
which has the drawback not to be consistent with asymptotic freedom as found in QCD and has motivated our improvement in Sec.~\ref{modification}. Note that the parameter $\kappa_{\textrm{\tiny UV}}$ enters in the expansion of $\beta(\alpha)$, Eq.~(\ref{eq:betahPirner}). A well defined function $\beta(\alpha)$ from perturbation theory doesn't have this parameter, and this is the starting point of the program followed in Refs.~\cite{Gursoy:2007cb,Gursoy:2007er,Gursoy:2008za}. As we explain in Sec.~\ref{sec:soldilaton} we fix $\kappa_{\textrm {\tiny UV}}$ and the scale parameter $\omega_{\textrm {\tiny UV}}$ (equivalently  $\kappa_{\textrm {\tiny IR}}$ and $\omega_{\textrm {\tiny IR}}$) by using two input values for $\alpha$, cf. Eq.~(\ref{runningcouplingvalues}).

\vspace{1cm}

\appendix{\textsf{\large Appendix B: Infrared Properties of the Improved Gravity Dual Theory}}
\label{sec:apB}

In this appendix we study the infrared properties of the 5-dim Nambu-Goto theory with the improved metric ${\bar h}(z)/(\Lambda z)^2$, cf. Eq.~(\ref{eq:metricbar}), which is proposed in Sec.~\ref{modification} and further developed in Secs.~\ref{calculation_after_modification} and \ref{sec:dilaton-runningcoupling}. The ultraviolet properties are fully dictated by the UV behavior of the $\beta$-function, so we will just focus on the IR asymptotics. As we will see later, the warp factor ${\bar h}(z)$ has a singularity at some finite value ${\bar z}_{\textrm {\tiny IR}}$, so that the coordinate $z$ is restricted to $z < {\bar z}_{\textrm {\tiny IR}}$, like in the guessed metric $h(z)/(\Lambda z)^2$ studied in Appendix~A. 

The IR expansion of $A(z)$ can be computed from the function $A(\alpha)$ and the functional dependence $\alpha(z)$. In the IR, i.e. $\alpha \to \infty$, Eq.~(\ref{eq:Aalpha}) can be integrated out using the expression of the $\beta$-function given in Eq.~(\ref{newbeta}). One gets

\begin{equation}
A(\alpha) = A(\alpha_*) - \frac{1}{b_2} \log\left( \frac{\alpha}{\alpha_*}  \right) + \dots \,  \qquad \alpha\to\infty \,. \label{eq:Aapp}
\end{equation}

The computation of $e^D$ involves an integration from $0$ to $\alpha$, and so it must be performed more carefully in order to retain the UV convergence, cf. Eq.~(\ref{eq:eDDD}). Inserting the full expression of $\beta(\alpha)$ into the integrand of Eq.~(\ref{eq:eDDD}), and considering an expansion at large $\alpha$, one gets

\begin{eqnarray}
\frac{4}{3}\int_0^\alpha \frac{\beta(a)}{3 a^2} da &=& -\frac{4}{9} b_2 \log \left(\frac{\alpha}{\bar\alpha} \right) + C_0  + {\cal O}(\alpha^{-4}) \nonumber \\
&&-\frac{2}{9} e^{-\frac{\alpha}{\bar\alpha}} \left[  \left(b_2-2\bar\alpha(b_0 + b_1 \bar\alpha)\right)\frac{\alpha}{\bar\alpha} + {\cal O}(\alpha^0) \right] \,, \qquad \alpha\to\infty \,,  \label{eq:intbetaApb}
\end{eqnarray}
where for simplicity we have defined the constant
\begin{equation}
C_0 = \frac{2}{9} \left[  b_2 \left( 3-2 \gamma_E \right) -2 \bar\alpha (2 b_0 + b_1 \bar\alpha) \right] \,.
\end{equation}
The integration is convergent in the UV, and the dominant contribution in the IR comes from the logarithmic term in the r.h.s. of Eq.~(\ref{eq:intbetaApb}). From Eq.~(\ref{eq:eDDD}), and using Eqs.~(\ref{newbeta}) and (\ref{eq:intbetaApb}) one can easily compute the IR asymptotics of $e^D$. It reads

\begin{equation}
e^D = \frac{e^{C_0}}{b_2 \alpha} \left( \frac{\alpha}{\bar\alpha}  \right)^{-\frac{4}{9}b_2}  \left[ 1 + {\cal O}(e^{-\frac{\alpha}{\bar\alpha}})  \right] \,, \qquad \alpha\to\infty \,. \label{eq:eDapp}
\end{equation}

The functional dependence $\alpha(z)$ can be computed from Eq.~(\ref{eq:dadz1}), which one can write in the following way

\begin{equation}
dz = {\bar\ell} \, e^{D-A} d\alpha \,. \label{eq:dzda1}
\end{equation}
Then the function $z(\alpha)$ follows from an integration of Eq.~(\ref{eq:dzda1}),

\begin{equation}
\int_{z_1}^z dz = { \bar\ell } \int_{\alpha_1}^{\alpha(z)} e^{D(a)-A(a)} da \,, \qquad \alpha_1 = \alpha(z_1) \,. \label{eq:intza}
\end{equation}
Inserting Eqs.~(\ref{eq:Aapp}) and (\ref{eq:eDapp}) into the r.h.s. of Eq.~(\ref{eq:intza}), we can perform the integration analytically. After inversing the solution, one gets

\begin{equation}
\alpha(z) = \left[ C_* (z_1 - z) + \alpha_1^{\frac{1}{b_2} - \frac{4}{9} b_2} \right]^{-\frac{b_2}{\frac{4}{9} b_2^2-1}} + \dots = \frac{1}{\left[C_* \cdot ({\bar z}_{\textrm{\tiny IR}}  - z)\right]^{\delta \cdot b_2} } + \dots   \,, \qquad  z \to {\bar z}_{\textrm{\tiny IR}} \,, \label{eq:azappB}
\end{equation}
where we have defined
\begin{eqnarray}
\delta = \frac{1}{\frac{4}{9}b_2^2-1}  \,, \qquad {\bar z}_{\textrm {\tiny IR}} = z_1 + \frac{\alpha_1^{\frac{1}{b_2}-\frac{4}{9}b_2}}{C_*} \,. \label{eq:deltazir}
\end{eqnarray}
and the constant $C_*$ is
\begin{equation}
C_* = \frac{1}{{\bar\ell} \cdot \delta}e^{A(\alpha_*) - C_0} \frac{\alpha_*^{\frac{1}{b_2}}}{{\bar\alpha^{\frac{4}{9}b_2}}}  \,.
\end{equation}

Inserting Eq.~(\ref{eq:azappB}) into Eq.~(\ref{eq:Aapp}) then one finally gets the IR asymptotics of $A(z)$, which reads

\begin{equation}
A(z) = \delta \cdot \log ( {\bar z}_{\textrm {\tiny IR}} - z) + \dots  \,, \qquad z \to {\bar z}_{\textrm{\tiny IR}} \label{eq:Azasyappa}  \,.
\end{equation}

The IR behavior of $\phi(z)$ can be obtained from Eqs.~(\ref{EEOM2}) and (\ref{eq:Azasyappa}), and it reads

\begin{equation}
\phi(z) =  -\frac{3}{2} \sqrt{\delta (1+\delta)} \log ( {\bar z}_{\textrm {\tiny IR}} - z  ) + \dots \,, \qquad z \to {\bar z}_{\textrm {\tiny IR}} \,,
\end{equation}
which corresponds to the asymptotics of $\log \alpha(z)$, cf. Eq.~(\ref{eq:azappB}), after taking into account Eq.~(\ref{eq:deltazir}).

The procedure to compute the IR asymptotics of the superpotential $W$ and the dilaton potential $V$ is explained in Appendix~A. The superpotential reads

\begin{equation}
W(\alpha) = \frac{9}{4} \delta \cdot C_*^{1+\delta} \cdot \alpha^{\frac{4}{9} b_2} + \dots   \,, \qquad \alpha\to\infty \,,
\end{equation}

and the dilaton potential

\begin{equation}
V(\alpha) = \frac{4}{3} C_*^{2+2\delta} \cdot \delta^2 \cdot (b_2^2-9) \alpha^{\frac{8}{9} b_2}  + \dots \,, \qquad \alpha\to\infty \,. \label{eq:VappB}
\end{equation}

Within the notation of Refs.~\cite{Gursoy:2007cb,Gursoy:2007er}, cf. Eq.~(\ref{eq:Vgursoy}), the solution of Eq.~(\ref{eq:VappB}) corresponds to $Q = \frac{4}{9}b_2$. As it has been discussed in Ref.~\cite{Gursoy:2007er}, the asymptotics of Eq.~(\ref{eq:Aapp}) leads to a confining theory whenever $Q > 2/3$, and this is fulfilled in our case for $b_2 > 3/2$. The IR behavior of $X(\alpha)$ reads

\begin{equation}
X(\alpha) = -\frac{1}{3} b_2 + \dots  \,, \qquad \alpha\to\infty 
\end{equation}  
so that the limit

\begin{equation}
\lim_{\alpha\to\infty} \left(  X(\alpha) + \frac{1}{2}\right) \log\alpha = K \,,
\end{equation} 
leads to $K=-\infty$ in the confining case, i.e. $b_2 > 3/2$, as it was explained in Sec.~\ref{modification}, cf. Eq.~(\ref{eq:critconfbeta}).

Giving the IR asymptotics of Eq.~(\ref{eq:Vgursoy}), a good singularity  according to the criterion of Ref.~\cite{Gursoy:2007er} is obtained when $Q < 2 \sqrt{2}/3$. This means that our theory is confining and it presents a good singularity for values of $b_2$ in the range

\begin{equation}
\frac{3}{2} < b_2 < \frac{3\sqrt{2}}{2}  \,.
\end{equation}
Nevertheless, as we will see below the upper bound is too conservative, and a slightly larger value is obtained when computing the IR asymptotics of the effective Schr\"odinger potential for the glueball spectrum. To this end, first we will analyze the $2^{++}$ sector, whose potential~is

\begin{equation}
V_2^{\textrm{Schr.}} (z) =  (B_2^\prime(z))^2 + B_2^{\prime\prime}(z) \,,
\end{equation}
where
\begin{equation}
B_2(z) = \frac{3}{2} A(z) \,. \label{eq:Btappb}
\end{equation}
After inserting the IR asymptotics of $A(z)$, Eq.~(\ref{eq:Azasyappa}), into Eq.~(\ref{eq:Btappb}) and computing the derivatives, one gets the following asymptotics for the effective Schr\"odinger potential

\begin{equation}
V_2 (z) = \frac{3}{2}\delta \left( \frac{3}{2}\delta - 1 \right) \frac{1}{({\bar z}_{\textrm {\tiny IR}} - z)^2} + \dots \,, \qquad z \to {\bar z}_{\textrm {\tiny IR}} \,. \label{eq:Vtappb}
\end{equation}

The potential diverges to $+\infty$ whenever $\delta > 2/3$, or equivalently

\begin{equation}
b_2 < \frac{3}{2} \sqrt{\frac{5}{2}} \approx 2.37 \,. \label{eq:b2gs}
\end{equation}

Otherwise the Schr\"odinger potential would diverge to $-\infty$ at $z \to {\bar z}_{\textrm {\tiny IR}} $, so that the singularity would be attractive to physical fluctuations. So, the condition Eq.~(\ref{eq:b2gs}) ensures that the singularity is of the good king according to the criterion of G\"ursoy et al.~\cite{Gursoy:2007er}.

The computation in the $0^{++}$ sector is similar. In this case the Schr\"odinger potential reads

\begin{equation}
V_0 (z) = (B_0^\prime(z))^2 + B_0^{\prime\prime}(z) \,,
\end{equation} 
where
\begin{equation}
B_0(z) = \frac{3}{2} A(z) + \frac{1}{2}\log [ X^2(z) ]  \,, \label{eq:Bsappb}
\end{equation}
and

\begin{equation}
X(z) = \frac{\beta(\alpha(z))}{3 \alpha(z)} \,. \label{eq:Xappb}
\end{equation}

Using the asymptotics of $A(z)$, Eq.~(\ref{eq:Azasyappa}), in Eq.~(\ref{eq:Bsappb}) and the asymptotics of $\alpha(z)$, Eq.~(\ref{eq:azappB}), in Eq.~(\ref{eq:Xappb}), one gets for $V_0(z)$ the same IR asymptotics as for $V_2(z)$, cf. Eq.~(\ref{eq:Vtappb}).

\begin{figure}[!ht]
  \begin{center}
     \epsfxsize 10cm \epsffile{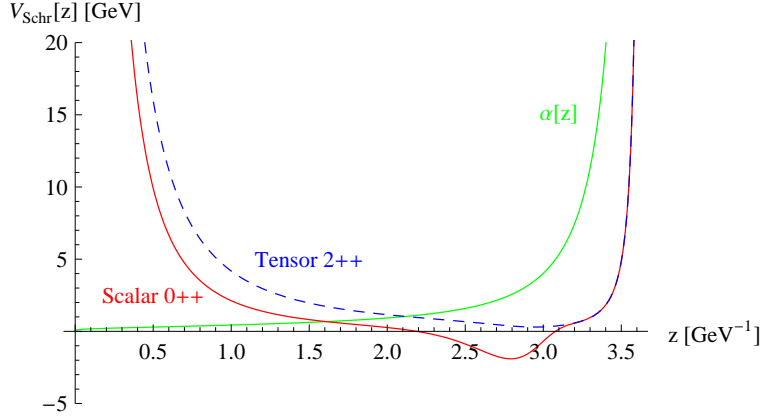}
  \end{center}
  \caption{\small The effective Schr\"odinger potentials for the $0^{++}$ (full red line) and $2^{++}$ (dashed blue line) glueball spectrum using $b_2 = 2.3$. We also show in full green line the running coupling. }
\label{fig:Vsch23}
\end{figure}

We plot in Fig.~\ref{fig:Vsch23} the effective Schr\"odinger potentials for the computation of the $0^{++}$ and $2^{++}$ glueball spectrum using the guessed value $b_2 = 2.3$.

\vspace{1cm}

\appendix{\textsf{\large Appendix C: Heavy $Q\bar{Q}$-Potential at Short Distance}}
\label{sec:apC}

In this appendix we give technical details for the analytical computation of the heavy quark-antiquark potential in the small separation limit as an expansion in powers of $\alpha_0$. The relevant formulas are Eqs.~(\ref{finalrho}) and (\ref{finalVqq}). The result up to NLO has been derived in Ref.~\cite{Zeng:2008sx}. 

In order to compute the series up to NNLO, we need to consider the QCD
$\beta$-function up to order $\alpha_0^3$, so we have to take into
account the parameters $b_0$ and $b_1$ (see Eq.~(\ref{QCDbeta})). We
have explicitly checked that the order $\alpha_0^4$ in the
$\beta$-function contributes at higher orders in the potential. Here
and in the following we use the notation $\alpha = \tilde\alpha
\cdot\alpha_0$. The variable $\tilde\alpha$ is in the interval $0 \le
\tilde\alpha \le 1$ in the integral of Eq.~(\ref{finalrho}) and the
first integral of Eq.~(\ref{finalVqq}).

First we will consider the computation of $\rho(\alpha_0)$ which is
given by Eq.~(\ref{finalrho}). The function $A(\alpha)$ can be
computed using Eq.~(\ref{eq:Aalpha}), and the result is
\begin{equation}
A(\alpha) = \frac{1}{b_0 \alpha} + \frac{b_1}{b_0^2} \log \left( \frac{\alpha}{b_0 + b_1 \alpha} \right) + {\cal O}(\alpha) \,. \label{eq:appAalpha}
\end{equation}
The function $e^D$ follows from Eq.~(\ref{eq:eDDD}), and after an expansion in powers of $\alpha_0$ it reads
\begin{equation}
e^D =  \frac{1}{b_0 \alpha^2} \left(
1 + d_1 \alpha_0 + d_2 \alpha_0^2 + {\cal O}(\alpha_0^3)
\right), \label{eq:appeD}
\end{equation}
where
\begin{eqnarray}
d_1 & = & -\left( \frac{4}{9} + \frac{b_1}{b_0^2} \right)b_0 \tilde\alpha\, , \\
d_2 & = & \frac{1}{81 b_0^2} \left(8 b_0^4 + 18 b_0^2 b_1 + 81 b_1^2 \right)
\tilde\alpha^2  \,.
\end{eqnarray}
 The integration in
$\tilde\alpha$ cannot be computed analytically even after this
expansion. In order to avoid this problem, we consider the following
series
\begin{equation}
\frac{e^{-3 \tilde{A}}}{\sqrt{1- \tilde\alpha^{-\frac{8}{3}}e^{-4\tilde{A}}}} =
\sum_{n=1}^\infty \frac{(2n)(2n-1)!!}{(2n-1)(2n)!!}
\tilde\alpha^{\frac{8}{3}(1-n)} e^{(1-4n)\tilde{A}} \,. \label{eq:appe3A}
\end{equation}
Note that $\tilde{A} = A(\alpha)-A(\alpha_0)$ is also a function of
the coupling constant $\alpha_0$, so it makes sense to consider an
expansion in $\alpha_0$ for the above expression. The result is
\begin{equation}
e^{(1-4n)\tilde{A}} = e^{(1-4n)\frac{1-\tilde\alpha}{b_0 \alpha_0 \tilde\alpha}}
\tilde\alpha^{(1-4n) \frac{b_1}{b_0^2}} \left[  
1+ t_1 \alpha_0 + t_2 \alpha_0^2 + {\cal O}(\alpha_0^3)
\right],  \label{eq:appenA}
\end{equation}
where
\begin{eqnarray}
t_1 & = & (1-4n)(1-\tilde\alpha) \frac{b_1^2}{b_0^3} \,,  \\
t_2 & = & -(1 - 4 n) (1 - \tilde\alpha) \frac{b_1^3}{2 b_0^6}\left[b_0^2(1 +
\tilde\alpha) - b_1 (1 - 4 n) (1 - \tilde\alpha)\right] \,. 
\end{eqnarray}
In Eq.~(\ref{eq:appenA}) there is a dependence in $\alpha_0$ that cannot be
expanded due to an essential singularity. Combining Eq.~(\ref{eq:appeD}) and
Eq.~(\ref{eq:appenA}), we get
\begin{equation}
e^D e^{(1-4n)\tilde{A}} = \frac{1}{b_0 \alpha_0^2} e^{(1-4n)\frac{1-\tilde\alpha}{b_0
\alpha_0 \tilde\alpha}} \tilde\alpha^{(1-4n) \frac{b_1}{b_0^2}-2} \left[ 1 + (d_1
+ t_1) \alpha_0 + (d_2 + d_1 t_1 + t_2)\alpha_0^2 + {\cal O}(\alpha_0^3)
\right] \,.
\end{equation}
Using this expression and taking into account Eq.~(\ref{finalrho}) and
Eq.~(\ref{eq:appe3A}), we can compute the integration in $\tilde\alpha$
analytically for every order in $\alpha_0$ and $n$, and the result is a
summation of terms involving the incomplete Gamma functions. Making a further
expansion in $\alpha_0$ one gets
\begin{equation}
\rho(\alpha_0) = 2 \bar\ell e^{-A_0} \left[  \rho_0 + \rho_1\alpha_0 + \rho_2
\alpha_0^2  + {\cal O}(\alpha_0^3)
 \right], \label{eq:rhoexpand1}
\end{equation}
where
\begin{eqnarray}
\rho_0 & = & \sum_{n=1}^\infty \frac{(2n)(2n-1)!!}{(2n-1)(4n-1)(2n)!!} \approx 0.596
\,, \nonumber \\
\rho_1 & = & \frac{16}{9}b_0 \sum_{n=1}^\infty
\frac{n(n-1)(2n-1)!!}{(2n-1)(4n-1)^2(2n)!!} \approx 0.0471 b_0 \,, \label{eq:rhos} \\
\rho_2 & = & \sum_{n=1}^\infty \frac{(4n)(2n-1)!!}{81(2n-1)(4n-1)^3 (2n)!!} 
\left[
4(22+n(40n-53))b_0^2 + 9(4n-1)(8n-5)b_1
\right] \,.\nonumber \\
&\approx&  0.0860 b_0^2 +0.180 b_1 \,. \nonumber
\end{eqnarray}
We truncate the summation of the infinite series to
$\sum_1^{20000} \cdots$, by which the residue contribution can be neglected to
$10^{-3}$ relative precision.

Next, we will focus on the computation of $V_{Q\bar{Q}} $, which is
given by Eq.~(\ref{finalVqq}). There are two contributions which we
call $V$ and $V_s$, corresponding to the first and second integrals
respectively, so we write $V_{Q\bar{Q}} = V - V_s$. Note that for the
first one, the expression is similar to that of $\rho(\alpha_0)$, and
so we can apply the same procedure that we explained above. In order to
make the integration analytical we consider the series
\begin{equation}
\frac{e^{\tilde{A}}\left(1-\sqrt{1-\tilde\alpha^{-\frac{8}{3}}
e^{-4\tilde{A}}}\right)}{\sqrt{1-\tilde\alpha^{-\frac{8}{3}}
e^{-4\tilde{A}}}} = \sum_{n=1}^\infty \frac{(2n-1)!!}{(2n)!!}
\tilde\alpha^{-\frac{8}{3}n} e^{(1-4n)\tilde{A}} \,. \label{eq:appeA}
\end{equation}
Using Eqs.~(\ref{finalVqq}), (\ref{eq:appeD}), (\ref{eq:appenA}) and
(\ref{eq:appeA}) we can compute analytically the integration in
$\tilde\alpha$ in the same way as we did for $\rho(\alpha_0)$. Finally
we arrive at this result
\begin{equation}
V(\alpha_0) = \frac{\bar\ell \alpha_0^{\frac{4}{3}}e^{A_0}}{\pi \bar l_s^2} \left[
v_0 + v_1 \alpha_0 + v_2 \alpha_0^2 + {\cal O}(\alpha_0^3)
\right] \,, \label{eq:appVresult}
\end{equation}
where
\begin{eqnarray}
v_0 & = & \sum_{n=1}^\infty \frac{(2n-1)!!}{(4n-1)(2n)!!} \approx 0.398 \,,\nonumber \\
v_1 & = & \frac{8}{9} b_0 \sum_{n=1}^\infty
\frac{(n-1)(2n-1)!!}{(4n-1)^2(2n)!!} \approx 0.0423 b_0\,, \label{eq:Vcoeff} \\
v_2 & = & \frac{2}{81}\sum_{n=1}^\infty \frac{(2n-1)!!}{(4n-1)^3 (2n)!!}
\left[
4(22+n(40n-53))b_0^2 + 9(4n-1)(8n-5)b_1
\right] \nonumber \\
& \approx & 0.0640 b_0^2 + 0.131 b_1 \,. \nonumber 
\end{eqnarray}
The computation of $V_s(\alpha_0)$ is a little bit tricky because it
does not allow similar expansion techniques. First of all it is
convenient to change the integration limits from $(\alpha_0,\infty)$
to $(0,1)$, and so we consider the variable replacement $\alpha
\rightarrow \frac{\alpha_0}{\hat\alpha} $. Then we get
\begin{equation}
V_s =  \frac{\bar\ell \alpha_0^{\frac{4}{3}}e^{A_0}}{\pi \bar l_s^2} 
\int_{\alpha_0}^\infty \tilde{\alpha}^{\frac{4}{3}}\cdot e^{D+\tilde{A}} d\alpha
= \frac{\bar\ell \alpha_0^{\frac{7}{3}}e^{A_0}}{\pi \bar l_s^2} \int_{0}^1
\hat{\alpha}^{-\frac{10}{3}}\cdot e^{D+\tilde{A}} d\hat\alpha =: \frac{\bar\ell
e^{A_0}}{\pi \bar l_s^2} \int_{0}^1 \hat{\alpha}^{-\frac{10}{3}}
f(\hat\alpha,\alpha_0) d\hat\alpha \,, \label{eq:appVs}
\end{equation}
where we have defined the function
\begin{equation}
f(\hat \alpha,\alpha_0) = \alpha_0^\frac{7}{3} e^{D+\tilde{A}} \,, \label{eq:f_def}
\end{equation}
which involves all the interesting dependence in $\alpha_0$. We can expand this
function in powers of $\alpha_0$ by considering its derivatives
\begin{equation}
f(\hat\alpha,\alpha_0) = f(\hat\alpha,0) + f^\prime(\hat\alpha,0) \alpha_0 +
\frac{1}{2} f^{\prime\prime}(\hat\alpha,0) \alpha_0^2 + {\cal O}(\alpha_0^3) \,,
\label{eq:f_series}
\end{equation}
where $f^\prime$ stands for derivative with respect to the second argument. This expansion can be easily done analytically, but the integration in $\hat\alpha$ must be computed numerically. By inserting Eq.~(\ref{eq:f_series}) into
Eq.~(\ref{eq:appVs}),  and after making the integration, finally we get
\begin{equation}
V_s(\alpha_0) =  \frac{\bar\ell \alpha_0^{\frac{4}{3}}e^{A_0}}{\pi \bar l_s^2} \Big[
1 + 0.889 b_0 \alpha_0 + (2.17 b_0^2 + 1.11 b_1)\alpha_0^2 +  {\cal
O}(\alpha_0^3)
\Big] \,. \label{eq:appVsresult}
\end{equation} 
Note that the contribution of the substraction term is larger than Eq.~(\ref{eq:appVresult}) by approximately a factor $3$ in the regime which we are interested in. Eq.~(\ref{eq:appVsresult}) reproduces well the full numerical computation of Eq.~(\ref{eq:appVs}) up to $\alpha_0 \simeq 0.19$, with an error of $5\%$. This $\alpha_0$ corresponds to distance $\rho \simeq 0.20 \,\textrm{GeV}^{-1}$.

Combining Eqs.~(\ref{eq:appVresult}), (\ref{eq:Vcoeff}) and (\ref{eq:appVsresult}), we obtain for the ${Q\bar{Q}}$-potential
\begin{equation}
  V_{Q\bar{Q}}(\rho) = V - V_s = -\frac{\bar\ell \alpha_0^{\frac{4}{3}}e^{A_0}}{\pi
\bar l_s^2}  \Big[0.602 +0.847 b_0 \alpha_0 + (2.106 b_0^2 + 0.979 b_1)\alpha_0^2 +
{\cal O}(\alpha_0^3)  \Big] \,. \label{eq:appVqqresult}
\end{equation}
From Eq.~(\ref{eq:rhoexpand1}) and Eq.~(\ref{eq:appVqqresult}) we can compute the
dimensionless quantity $\rho V_{Q\bar{Q}}(\rho)$ up to order $\alpha_0^2$, and
we get the final result
\begin{equation}
V_{Q\bar{Q}}(\rho) = -\frac{2 \bar\ell^2}{\pi \bar l_s^2}
\frac{\alpha_0^{4/3}(\rho)}{\rho}
\Big[
0.359 + 0.533 b_0 \alpha_0(\rho) + (1.347 b_0^2 + 0.692 b_1) \alpha_0^2(\rho) +
{\cal O}(\alpha_0^3)
\Big]   \,. \label{eq:appValpha1}
\end{equation}
The function $\alpha_0(\rho)$ can be obtained by reversing
Eq.~(\ref{eq:rhoexpand1}). To do that we make first an expansion of $A_0$ in $\alpha_0$. The resulting expression is
\begin{equation}
\alpha_0(\rho) = \frac{1}{b_0 \log\left(\frac{c}{\rho}\right) -\frac{b_1}{b_0} \log \alpha_0(\rho)} 
- \frac{\left( b_0 \frac{\rho_1}{\rho_0} +\frac{b_1^2}{b_0^2}\right) \alpha_0}{\left(b_0 \log\left(\frac{c}{\rho}\right) -\frac{b_1}{b_0} \log \alpha_0(\rho) \right)^2}  + {\cal O}\left( \frac{\alpha_0^2}{\log^2 \left(\frac{c}{\rho}\right)} \right), \label{eq:apprhosmall}
\end{equation}
where
\begin{equation}
c \equiv 2 \rho_0 b_0^{b_1/b_0^2} \bar\ell \approx 1.32 \bar\ell \,, \label{eq:Appc}
\end{equation}
with the parameters $\rho_0$ and $\rho_1$ defined in Eq.~(\ref{eq:rhos}). Then we can use an iteration method to obtain the solution order by order. For example, the first order is given by $\left(b_0\log\left(\frac{c}{\rho} \right)\right)^{-1}$. If one substitutes this expression in the r.h.s. of Eq.~(\ref{eq:apprhosmall}), then we get the NLO approximation.

Combining Eq.~(\ref{eq:appValpha1}) and Eq.~(\ref{eq:apprhosmall}) we have the
$Q\bar{Q}$-potential as a function of the separation $\rho$ between the quark
and the antiquark in the limit of small separations, i.e. $\rho \rightarrow 0$. This result agrees with Ref.~\cite{Zeng:2008sx} at leading order. We corrected a mistake in this reference at next to leading order for $V_s$. Fully NNLO expressions are provided here.

\bibliographystyle{h-physrev3}
\bibliography{phenads}

\end{document}